\documentclass[fleqn,usenatbib]{mnras}  
\usepackage{newtxtext,newtxmath}
\usepackage{xfrac}

\usepackage{ae,aecompl}

\usepackage{graphicx}	
\usepackage{amsmath}	
\usepackage[dvipsnames]{xcolor} 
\usepackage{microtype}
\usepackage[outline]{contour}
\usepackage{tikz}
\usepackage{algorithm}
\usepackage{algpseudocode}
\usepackage{bbold}

\usepackage{subcaption}

\newcommand{\be}{\begin{equation}}
\newcommand{\ee}{\end{equation}}
\newcommand{\fett}[1]{\boldsymbol{#1}}
\newcommand{\dd}{{\rm d}}

\newcommand{\rb}{{\rm b}}

\newcommand{\rM}{{\rm m}}

\newcommand{\ini}{{\rm ini}}

\newcommand{\nab}{\fett{\nabla}}
\newcommand{\Mpc}{{\rm Mpc}}
\newcommand{\Mpch}{\,{\rm Mpc}\, h^{-1}}

\newcommand{\nabx}{\fett{\nabla}_{\fett{x}}}
\newcommand{\nabL}{\fett{\nabla}^{\rm L}}

\algnewcommand\algorithmicto{\textbf{to}}
\algrenewtext{For}[3]%
  {\algorithmicfor\ #1 $\gets$ #2 \algorithmicto\ #3 \algorithmicdo}

\definecolor{lime}{HTML}{A6CE39}
\DeclareRobustCommand{\orcidicon}{
	\begin{tikzpicture}
	\draw[lime, fill=lime] (0,0) 
	circle [radius=0.14] 
	node[white] {{\fontfamily{qag}\selectfont \tiny ID}};
	\draw[white, fill=white] (-0.0625,0.095) 
	circle [radius=0.007];
	\end{tikzpicture}
	\hspace{-2mm}
}

\foreach \x in {A, ..., Z}{%
	\expandafter\xdef\csname orcid\x\endcsname{\noexpand\href{https://orcid.org/\csname orcidauthor\x\endcsname}{\noexpand\orcidicon}}
}

\contourlength{0.2pt} 
\contournumber{2} 

\title[Shell-crossing in $\mathit{\Lambda}$CDM]{Shell-crossing in a \contour{black}{$\Lambda$}CDM Universe}

\author[Rampf \& Hahn]{Cornelius Rampf$^{\,{\text{\tiny\orcidA{}}}\,\,\hyperlink{OCA}{1}}$\thanks{\!\!Marie Sk\l odowska--Curie Fellow; e-mail: \href{mailto:cornelius.rampf@oca.eu}{cornelius.rampf@oca.eu}} and Oliver Hahn$^{{\text{\tiny\orcidC{}}}\,\,\hyperlink{OCA}{1}, \hyperlink{AstroVienna}{2}, \hyperlink{MathVienna}{3}}$\thanks{\!E-mail: \href{mailto:oliver.hahn@univie.ac.at}{oliver.hahn@univie.ac.at}}  \hypertarget{OCA} \\
$^{1}$Universit\'e C\^ote d'Azur, Observatoire de la C\^ote d'Azur, CNRS, Laboratoire Lagrange, Boulevard de l'Observatoire, CS 34229, 06304 Nice, France  \hypertarget{AstroVienna} \\
$^{2}$Department of Astrophysics, University of Vienna, T\"urkenschanzstraße 17, 1180 Vienna, Austria  \hypertarget{MathVienna}\\
$^{3}$Department of Mathematics, University of Vienna, Oskar-Morgenstern-Platz 1, 1090 Vienna, Austria}

\date{Accepted XXX. Received YYY; in original form ZZZ}

\pubyear{2020}

\begin{document}
\label{firstpage}
\pagerange{\pageref{firstpage}--\pageref{lastpage}}
\maketitle

\begin{abstract}
Perturbation theory is an indispensable tool for studying the cosmic large-scale structure, and establishing its limits is therefore of utmost importance. One crucial limitation of perturbation theory is shell-crossing, which is the instance when cold-dark-matter trajectories intersect for the first time. We investigate Lagrangian perturbation theory (LPT) at very high orders in the vicinity of the first shell-crossing for random initial data in a realistic three-dimensional Universe. For this we have numerically implemented the all-order recursion relations for the matter trajectories, from which the convergence of the LPT series at shell-crossing is established. Convergence studies performed up to the 40th order reveal the nature of the convergence-limiting singularities. These singularities are not the well-known density singularities at shell-crossing but occur at later times when LPT already ceased to provide physically meaningful results. 
\end{abstract}


\section{Introduction}


\nocite{Uhlemann:2019,2017ApJ...837..181F,Rampf:2015mza,Michaux:2020,PODVIGINA2016320,MercerRoberts1990,Bardeen1986,vanDyke1974}

A given function $\Psi(D)$ that is analytic around $D=0$  can be locally represented by a convergent Taylor series.
In physical applications,
the  maximal value of the (temporal) variable~$D=D_\star$ for which the series converges is frequently unknown, always limited by the nearest singularity(ies) in the complex $D$-plane (sometimes at $D_\star = \infty$), and can be determined, e.g., by using d'Alembert's ratio test.

{\it Mutatis mutandis,} the same applies for the cosmological gravitational dynamics of cold dark matter (CDM), at least when formulated in Lagrangian coordinates. Indeed,  CDM trajectories can be represented by a convergent Taylor series \citep{Zheligovsky:2014,Rampf:2015mza}, however so far only rough estimates on the radius of convergence exist \citep{Michaux:2020}. 
The  framework for calculating the Taylor coefficients  is called Lagrangian perturbation theory (LPT), with pioneering work done by e.g.\ \cite{Zeldovich:1970,Buchert:1989xx,Buchert:1993xz,1992ApJ...394L...5B,Bouchet:1995,Ehlers:1996wg}. 
Nonetheless, without knowledge of the radius of convergence, spanned by $D_\star$, it is unclear until which time these solutions remain {\it mathematically} meaningful.

Furthermore, since standard LPT is based on a perfect pressureless fluid description, its predictions are no longer {\it physically} meaningful once matter trajectories have crossed for the first time. This instance is usually called shell-crossing (sc) and is accompanied by (formally) infinite densities.
Shell-crossing is well understood for simplified initial conditions \citep{Novikov:2010ta,2004ApJS..151..185Y,McQuinnWhite2016,Rampf:2017jan,Rampf:2017tne,Saga:2018,Taruya:2017,Pietroni:2018,Rampf:2019nvl}, but not so for realistic initial conditions. 

Curiously, an infinite density at shell-crossing is not necessarily a convergence-limiting singularity, since  in Lagrangian coordinates  the density is not a dynamical field but merely a derived quantity. It is the purpose of this work to investigate the rather distinct problems of finding the radius of convergence of LPT, and of detecting the first shell-crossing in a $\Lambda$CDM Universe with cosmological constant~$\Lambda$. We find that LPT converges fairly  fast, precisely since the radius of convergence surpasses the instance of the first shell-crossing.

\section{Setup}

We define the peculiar velocity of matter with $\fett{v} = a \partial_t \fett{x}$, where $\fett{x}=\fett{r}/a$
are comoving coordinates and $a$ is the cosmic scale factor. 
The basic equations that describe the evolution of matter elements in a $\Lambda$CDM Universe are
\be \label{eq:EOMs}
  \frac{\dd^2 \fett{x}}{\dd t^2} + 2H \frac{\dd \fett{x}}{\dd t} = - \frac{1}{a^2} \nabx \phi \,, 
\qquad \nabx^2 \phi = 4 \uppi G \bar \rho \,a^2\, \delta(\fett{x}) \,,
\ee
where $H$ is the Hubble parameter, $\bar \rho$ the mean matter density of the Universe, and $\delta = (\rho - \bar \rho)/\bar \rho$ the density contrast.

It is standard to solve these equations by introducing the matter trajectory $\fett{q} \mapsto \fett{x}(\fett{q},t) = \fett{q} + \fett{\psi}(\fett{q},t)$ from initial position $\fett{q}$ to current position $\fett{x}$ at time $t$. 
Here, $\fett{\psi}(\fett{q},t)$ is the Lagrangian displacement field  which plays a central role in solving the above equations in~LPT.
Until shell-crossing, mass conservation is 
\be \label{eq:density}
   \delta(\fett{x}(\fett{q},t)) = \frac{1}{J(\fett{x}(\fett{q},t))} -1 \,,
\ee
where $J = \det ( \nabL \fett{x} )$ is called the Jacobian, 
and $\nab^{\rm L}$ denotes a partial derivative in Lagrangian space. 
The first shell-crossing is achieved for $J(\fett{q}_{\rm sc},t_{\rm sc})\equiv 0$ at the smallest possible time $t_{\rm sc}$ and location $\fett{q}_{\rm sc}$. As it is well-known, at those locations, we have $\delta \to \infty$.

In standard Lagrangian perturbation theory,
Eqs.\,\eqref{eq:EOMs}--\eqref{eq:density} are solved with the following {\it Ansatz} for the displacement $\fett{\psi}= \fett{x}-\fett{q}$,
\be \label{eq:psiansatz}
  \fett{\psi}(\fett{q},t) = \sum_{n=1}^\infty \fett{\psi}^{(n)}(\fett{q})\,D^n \,,
\ee
where $D = D_+(a)$ is the linear growth function in $\Lambda$CDM which we use here as a refined time variable. Taking divergence and curl operations of~\eqref{eq:EOMs} and matching the powers in $D^n$, one obtains all-order recurrence relations for the spatial 
coefficients \citep{Rampf:2012b,Zheligovsky:2014,Rampf:2015mza,Matsubara:2015ipa}
\begin{subequations} \label{eq:recs}
\begin{align}
  &\nabL \cdot \fett{\psi}^{(n)} = - \varphi_{,ll}^\ini \delta_1^n + \sum_{0<s<n} \frac{(3-n)/2-s^2 -(n-s)^2}{(n+3/2)\,(n-1)} \mu_2^{(s,n-s)} \nonumber \\
 &\quad\,\,\,\, \qquad + \! \sum_{n_1+n_2+n_3=n} \! \frac{(3-n)/2 - n_1^2-n_2^2-n_3^2}{(n+3/2)\,(n-1)} \mu_3^{(n_1,n_2,n_3)} ,  \label{eq:scalarrec}  \\
 &\nabL \times \fett{\psi}^{(n)} = \frac 1 2 \sum_{0<s<n} \frac{n-2s}{n} \nabL \psi_k^{(s)} \times \nabL \psi_k^{(n-s)} \,, \label{eq:cauchy}
\end{align} 
\end{subequations}
where ``$,l$'' denotes a partial derivative with respect to component $q_l$, summation over repeated indices is assumed, and $\varphi^\ini$ is the initial gravitational potential provided at $a=0$ (suitably rescaled); see e.g.\ \cite{Michaux:2020} for details how $\varphi^\ini$ is obtained. Furthermore, $\mu_2^{(n_1,n_2)}=(1/2)[\psi_{l,l}^{(n_1)}\psi_{m,m}^{(n_2)}- \psi_{l,m}^{(n_1)}\psi_{m,l}^{(n_2)}]$ and $\mu_3^{(n_1,n_2,n_3)} = (1/6) \varepsilon_{ikl} \varepsilon_{jmn} \psi_{i,j}^{(n_1)} \psi_{k,m}^{(n_2)} \psi_{l,n}^{(n_3)}$ are purely spatial kernels of the Jacobian, and $\varepsilon_{ijk}$ the Levi--Civit\`a symbol. 
Note that we ignore sub-dominant 
terms $\propto \Lambda$ in~Eqs.\,\eqref{eq:recs} since we are here interested in observables at rather high redshifts ($z\gg 5$) where they are vanishingly small; this could be rectified if needed.

\paragraph*{Summary of numerical implementation.}
The divergence and curl relations~\eqref{eq:recs} define a Helmholtz decomposition from which the displacement field can be determined to arbitrary high precision in perturbation theory. These recursion relations can be translated into an efficient algorithm in Fourier space, which we have implemented in the  publicly available software package {\sc Music2-MonofonIC}.\footnote{available from \url{https://bitbucket.org/ohahn/monofonic}.}
The numerical code is parallelized (MPI+threads), 
handles convolutions and space derivatives by employing fast Fourier transforms,
and is fully de-aliased. Regarding the latter, this turns out to be somewhat memory intensive due to the appearance of cubic terms in Eq.\,\eqref{eq:scalarrec} for which the computational domain needs to be temporarily extended by a factor twice as much as the usual Orszag's 3/2 rule \citep{Roberts11}, requiring eight times the memory. In addition, the $\fett{\psi}^{(n)}$'s need to be stored temporarily.  Thus, the primary limitation of generating the displacement on grid points is due to limited working memory.

For the cosmological parameters we set $\Omega_\rM  = 0.302$, $\Omega_\rb = 0.045$, $\Omega_\Lambda = 0.698$, $h = 0.703$, $\sigma_8 = 0.811$, $n_s = 0.961$. If not otherwise stated, we evaluate the fields at $N=256^3$ grid points with box length $L_{\rm box} = 125 \Mpch$ for which $k_{\rm Ny} = 6.43 h\, \Mpc^{-1}$ is the Nyquist frequency. 
LPT results are sensitive to the shape of the initial power spectrum, and to elucidate this we show results for standard CDM, as well as for a {\it fictitious} warm dark matter (WDM) model with an unrealistically large mass $m_{\rm wdm} = 0.25$\,keV normalized with respect to $\sigma_8$, where we employ the \cite{1998ApJ...496..605E} transfer functions. Details to the used algorithms and numerical tests are provided in the supplementary material~\ref{app:implementation}.

\section{Shell-crossing and radius of convergence}

After having numerically implemented perturbative solutions of the displacement to arbitrarily high order, it is natural to ask whether Eq.\,\eqref{eq:psiansatz} defines a convergent series? And if yes, what are the criteria for its breakdown?

Regarding the former, \cite{Zheligovsky:2014} and \cite{Rampf:2015mza} determined lower bounds on the radius of convergence, 
implying that the LPT series is time-analytic and converges at least until some finite time. However, these theoretical methods cannot predict the actual radius of convergence for random initial conditions, for which numerical tests should be employed -- as we do below.

Regarding the criteria for its breakdown, it is clear that, {\it physically}, Eq.\,\eqref{eq:psiansatz} remains only meaningful as long as the flow is (locally) in the single-stream regime. Indeed, once the first shell-crossing has occurred, some regions will be occupied with multiple fluid streams which gravitationally interact with each other. However, standard LPT does not encapsulate the resulting non-trivial accelerations in those multi-streaming regions. Instead of incorporating such effects, here we rather raise another, highly nontrivial question: for $\Lambda$CDM initial conditions, is the {\it actual} radius of convergence of the LPT series limited by the first instance of shell-crossing?

\begin{figure}
	\centering
	\includegraphics[width=0.98\columnwidth]{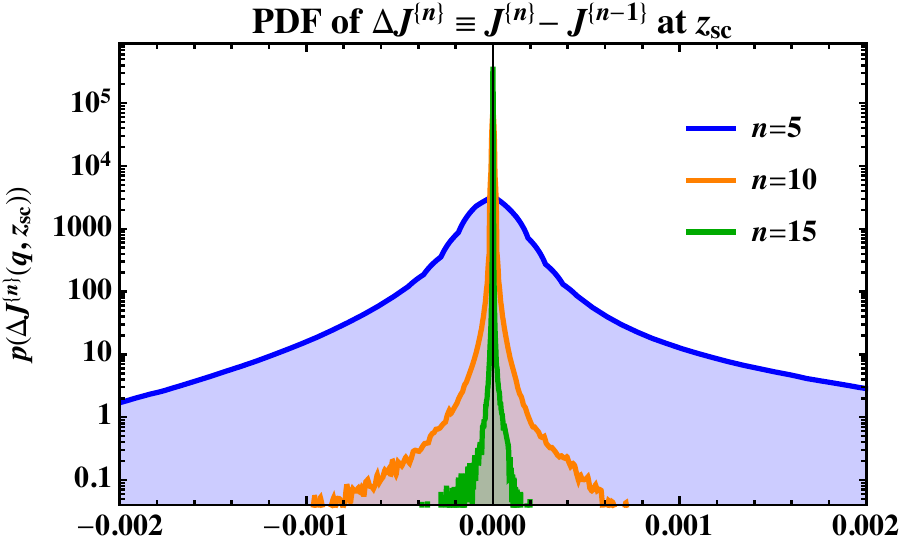}
	\caption{PDF of the residual $\Delta J^{\{n\}}(\fett{q}) = J^{\{n\}} -J^{\{n-1\}}$ for LPT orders
           $n=5,10,15$  (blue, orange and green lines), evaluated at the redshift of first shell-crossing, which for the present CDM case is at $z_{\rm sc} =14.77$. This figure demonstrates that LPT displays convergent behaviour at all grid locations. }
	\label{fig:diffJ}
\end{figure}

To dissect the distinct problems, we first study the convergence at the first shell-crossing. For this
we generate the displacement coefficients  at successively higher orders and search for the spatial grid point location $\fett{q}_{\rm sc}$ where the perturbatively truncated Jacobian 
\be
  J^{\{n\}} \equiv \det \bigg( \mathbb{1} + \nabL \sum_{s=1}^n \fett{\psi}^{(s)} (\fett{q})\, D^s \bigg)
\ee
 vanishes for the first time  ($\mathbb{1}$ is the unit matrix). 
This search determines $z_{\rm sc}^{(n)}$, which is the $n$th-order estimate of the redshift of shell-crossing. Then we iterate to higher orders and monitor the trend  of $z_{\rm sc}^{(n)}$ at successive orders. In all cases considered we find that $z_{\rm sc}^{(n)}$ converges to a stable answer, correct to two decimal places, starting at orders between $n=7-17$. Henceforth we call the converged redshift and corresponding shell-crossing location simply $z_{\rm sc}$ and~$\fett{q}_{\rm sc}$, respectively. Here the speed of convergence depends mostly on the topology of the seeds that collapse first. 
See section~\ref{sec:collapsestructures} for further details, and the supplementary material~\ref{app:zsc-results} for explicit case examples as well as sub-grid convergence studies.

Regarding the overall convergence of the Jacobian in the whole spatial domain, in Fig.\,\ref{fig:diffJ} we show the PDF resulting from taking the difference of perturbatively truncated Jacobians $\Delta J^{\{n\}}(\fett{q},z_{\rm sc}) = J^{\{n\}}(\fett{q},z_{\rm sc}) -J^{\{n-1\}}(\fett{q},z_{\rm sc})$ from all grid points at shell-crossing redshift (here: $z_{\rm sc}= 14.77$ for our base CDM model), for various perturbation orders $n$. As it is expected for a convergent series, we observe that the truncated distribution of $\Delta J^{\{n\}}$ approaches zero for increasing orders $n$. Furthermore, the tails in those PDFs lose their support for larger $n$'s: In the specific case shown in the figure, 
for $n=5,10,15$ the most extremal absolute deviations of $\Delta J^{\{n\}}$ from~0 are respectively 
$0.08144, 0.00916, 0.00157$.
Actually, even for those rare events with extremal deviations, which we frequently observed in void regions, LPT displays convergent behaviour until $z_{\rm sc}$ -- and even slightly beyond (by employing the Domb--Sykes method explained below). Thus, the Jacobian is converged to at least two decimal places at 10LPT at shell-crossing time -- and not just at~$\fett{q}_{\rm sc}$ but in the whole spatial domain.

We remark that the LPT convergence at $z_{\rm sc}$ can also be observed in Fourier space. For this we determine the power spectrum $P_J$ of the perturbatively truncated Jacobian at various redshifts, where $\langle \tilde J(\fett{k}_1)\, \tilde J(\fett{k}_2) \rangle = (2\uppi)^3 \delta_{\rm D}^{(3)} (\fett{k}_1+ \fett{k}_2) \,P_J(k_1)$, with a tilde denoting the Fourier transform,
$\delta_{\rm D}^{(3)}$  being the Dirac delta, and $k = || \fett{k} ||$. We find that, until the first shell-crossing, $P_J$ is already converged to sub per cent precision at 3LPT \citep[cf.][]{Michaux:2020}; for details we refer to the supplementary material~\ref{app:implementation} (see in particular Fig.\,\ref{fig:powerJ}).

\begin{figure}
	\centering
	\includegraphics[width=0.93\columnwidth]{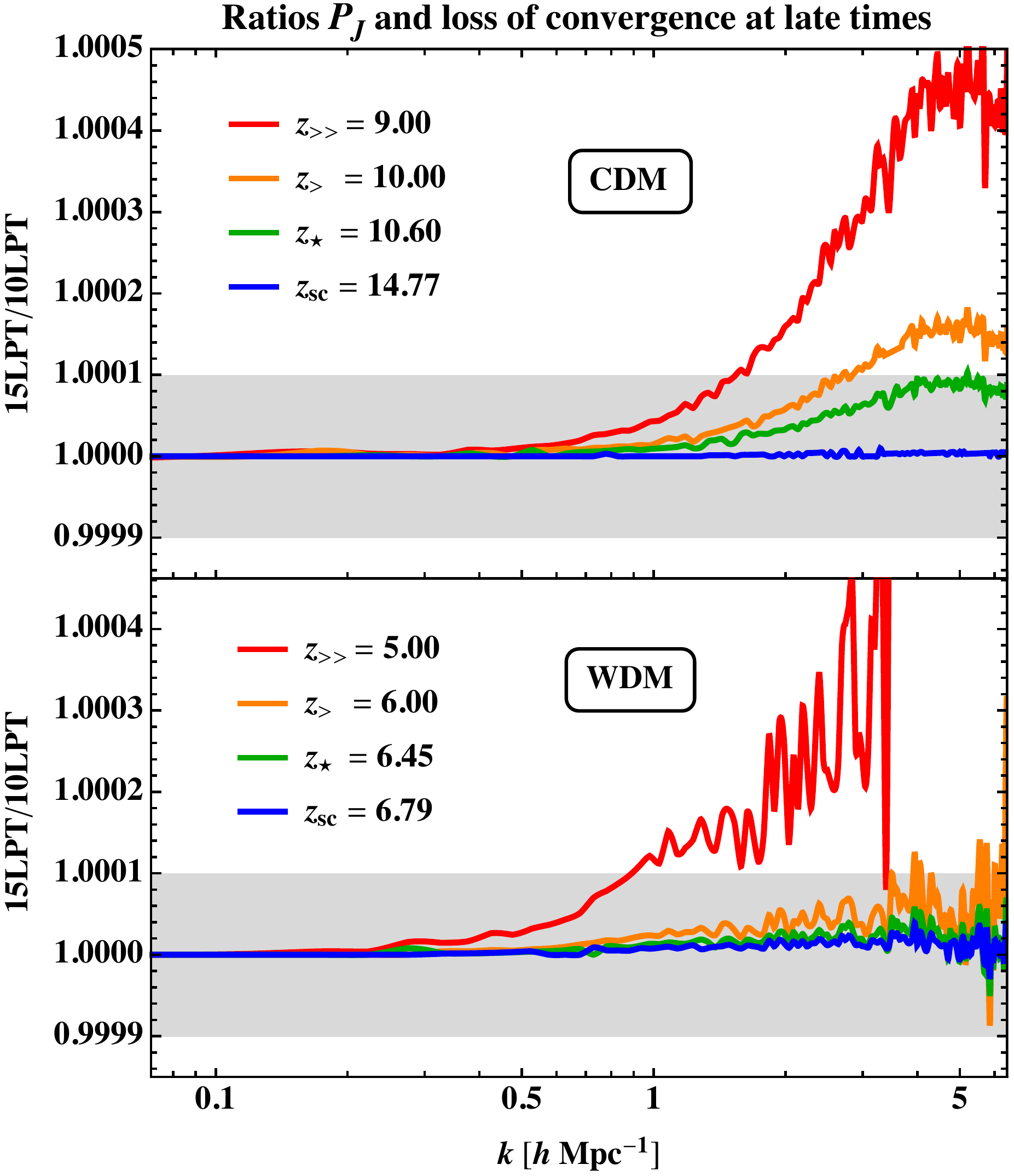}
	\caption{Ratio between power spectra of the truncated Jacobian $J^{\{n\}}$ for $n=15$ versus $n=10$ at various redshifts, for CDM (upper panel) and a fictitious WDM model with $m_{\rm wdm} = 0.25$\,keV (lower panel).  At the redshift of first shell-crossing ($z_{\rm sc}$), the ratios conform to unity to extremely high precision (grey shading denotes $10^{-4}$ deviation). Only at much later times, starting at $z_\star$ which marks the estimated redshift of convergence (see Fig.\,\ref{fig:dombsykes}),  the ratios begin to diverge, thereby revealing the loss of convergence. Evidently, the loss of convergence is rather subtle at those high redshifts, but gets vastly amplified later on (cf.\ Figs.\,\ref{fig:PJdivlpt} and~\ref{fig:ratioPJ40lpt}). Note that for results $z \leq z_{\rm sc}$, LPT does not incorporate multistreaming effects.}
	\label{fig:ratioJlatetime}
\end{figure}

Clearly, convergence is lost iff, for some later time and for sufficiently high perturbation orders, different refinement levels of perturbative truncations (e.g., for the Jacobian) begin to diverge. 
This is illustrated in Fig.\,\ref{fig:ratioJlatetime} where we show ratios of $P_J$ at 15th- versus 10th-order in LPT for various redshifts, for standard CDM (top panel) and a WDM model (bottom panel).
Evidently, the $10$th- and $15$th-order predictions for the Jacobian coincide to an extremely high precision at shell-crossing. Even more, this level of precision is maintained shortly after shell-crossing, denoted with~$z_\star$,  suggesting that the time of mathematical convergence of LPT surpasses the time of shell-crossing -- a conclusion that we have reached for all considered initial data and resolutions. Surpassing~$z_\star$, by contrast, generally leads to the loss of convergence which, shortly after $z_\star$, is rather subtle however amplifies and eventually  affects all relevant Fourier scales (see Figs.\,\ref{fig:PJdivlpt} and~\ref{fig:ratioPJ40lpt} for late-time results).  
We remark that, strictly speaking,  $J$ refers to the physical Jacobian only  until shell-crossing but not later; nonetheless, even at later times, $J$ remains a good indicator of the loss of mathematical convergence.

\begin{figure}
	\centering
	\includegraphics[width=0.88\columnwidth]{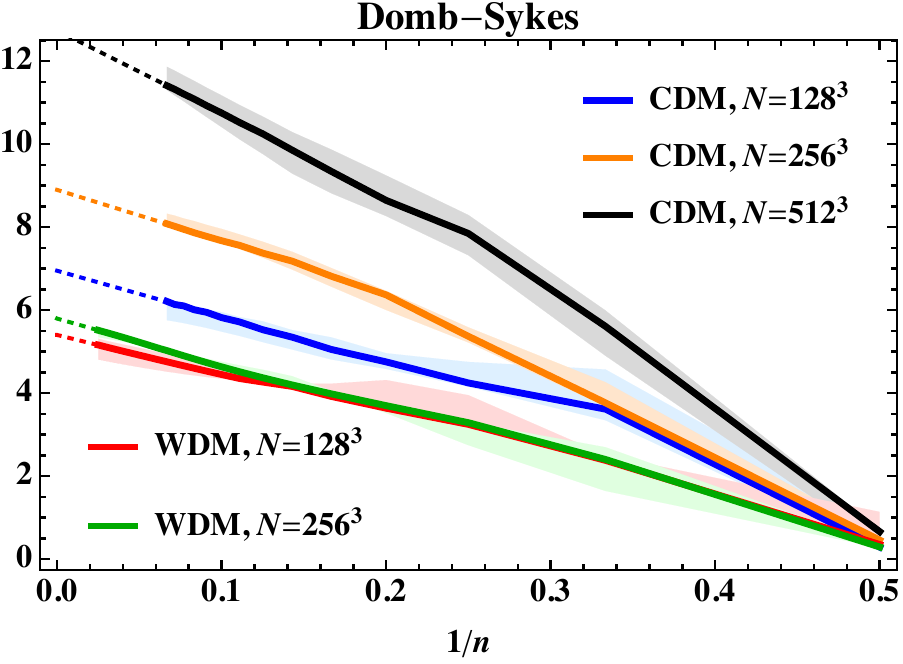}
	\caption{Domb--Sykes plot at first shell-crossing locations $\fett{q}_{\rm sc}$ for CDM [WDM] up to $n=15$ [$n=40$], drawn at various resolutions. Shown are the ratios of subsequent Taylor coefficients $\|\fett{\psi}^{(n)}\|/\|\fett{\psi}^{(n-1)}\|$ over $1/n$. For all data we set $L_{\rm  box}= 125\Mpch$. Lines denote the median obtained from 5 realizations, shaded regions its 32 and 68 percentiles, while the dotted lines are the respective linear extrapolations at large Taylor orders. The $y$ intercepts of those extrapolations reveal the numerical predictions of the inverse of the radius of convergence (see Table~\ref{tab:extrapol}).}
	\label{fig:dombsykes}
\end{figure}

To pin down the convergence rigorously, we consider the LPT series of the $L^2$ norms of displacement coefficients  
\be
  \Psi(\fett{q},t) \equiv \sum_{n=1}^\infty \left\| \fett{\psi}^{(n)}(\fett{q}) \right\| \,D^n \,,
\ee
and perform numerically the ratio test
\be \label{eq:ratiotest}
   \lim_{n \to \infty} \text{\LARGE $\sfrac{\text{\normalsize $\big\| \fett{\psi}^{(n)}  \big\|$}\,}{\,\text{\normalsize $\big\| \fett{\psi}^{(n-1)}\big\|$}}$}  = \frac{1}{D_\star} \,,
\ee
where $D_\star$ is the radius of convergence (when the limit exists). To do so,  we draw in Fig.\,\ref{fig:dombsykes} the Domb--Sykes plot \citep[cf.][]{DombSykes:1957} at the first shell-crossing location, averaged over five realizations. Specifically, for given realization we draw subsequent ratios of coefficients  $\|\fett{\psi}^{(n)}\|/\|\fett{\psi}^{(n-1)}\|$  versus $1/n$, and take the $y$-intercept (``$n \to \infty$'') as the estimate for $1/D_\star$ (or, conversely, the redshift of convergence $z_\star$). For WDM we go up to 40th order in LPT while for CDM we recommend not going beyond $n=15$ for reasons explained below. The ratios of displacement coefficients settle into a linear behaviour at sufficient high orders, which justifies  using a linear extrapolation to the $y$-intercept (denoted by dotted lines). 
In Table~\ref{tab:extrapol} we show the medians for $z_\star$ and $z_{\rm sc}$ together with their 32 and 68 percentile variances (for  $z_\star$ the variances result from linear extrapolations in the Domb--Sykes plot). For $z_{\rm sc}$ these variances span up the window when shell-crossing is most likely to occur.

\begin{table}
\begin{center}
\begin{tabular}{ccccccc}
 model  & $N$ & $m$        & $b$   & $\rho$              & \phantom{1}$z_\star$    & $z_{\rm sc}$  \\
\hline
 CDM        & $128^3$      & -11.16   & \phantom{1}6.95                & 0.61    & $\phantom{1}7.90_{-0.81}^{-0.21}$  &  $11.17_{-0.27}^{+0.15}$  \\[0.1cm]
 CDM        & $256^3$      & -12.30   & \phantom{1}8.90                & 0.38    & $10.40_{-0.19}^{+0.21}$            &  $15.57_{-0.80}^{+0.30}$ \\[0.1cm]
 CDM        & $512^3$      & -19.90   & 12.74                         & 0.56    &  $15.31_{+0.08}^{+0.55}$           &  $20.69_{-0.26}^{+0.27}$ \\[0.1cm]
 WDM        & $128^3$      & \phantom{1}-9.72   & \phantom{1}5.40      & 0.80    & $\phantom{1}5.91_{-0.54}^{+0.16}$  &  $\phantom{1}6.81_{-0.07}^{+0.13}$  \\[0.1cm]
 WDM        & $256^3$      & -11.26   & \phantom{1}5.80                & 0.94     & $\phantom{1}6.42_{-0.34}^{+0.06}$ &  $\phantom{1}6.96_{-0.12}^{+0.12}$ \\[0.1cm]
\hline
\end{tabular}
\end{center}
\caption{Results from the linear extrapolation  $y = m \,(1/n) + b$ of Taylor coefficients  $\|\fett{\psi}^{(n)}\|/\|\fett{\psi}^{(n-1)}\|$ at large orders $n$  (dotted lines in Fig.\,\ref{fig:dombsykes}). The estimated  singularity exponent is $\rho = -1 - m/b$,  while  the convergence redshifts are obtained from $z_\star=1/a(D)|_{D=D_\star}-1$, where $D_\star=1/b$.  Finally, $z_{\rm sc}$ denotes here the median of the first shell-crossing redshift,  and we have added the 32 and 68 percentiles as errors to $z_\star$ and $z_{\rm sc}$.}
\label{tab:extrapol}
\end{table}

Evidently the value for $z_\star$ is only mildly resolution dependent for WDM,  while for CDM the values of $z_\star$ differ substantially. This is due to the non-compact nature of the CDM spectrum, which leads to an intrinsic UV dependence at large~$k$. Nonetheless we expect that this UV dependence should become smaller when the Nyquist frequency is appropriately increased.  We leave such computationally challenging avenues for future work.

We note that the linear asymptotic behaviour of Taylor ratios is usually obtained also for single realizations. Only in rare cases we observe a mild oscillating behaviour superposed on a linear regression:  these oscillations appear to arise due to overlapping or adjacent singularities; see the supplementary material~\ref{app:outliers} for details. Nonetheless, even in such cases the ratios of Taylor coefficients still converge to a single point, and thus allow the accurate determination of $z_\star$  -- but collude identifying the nature of the singularities.

For the cases when the extrapolation of the Domb--Sykes plot is linear at large orders $n$, we can assume that the underlying singularities have a pole-like structure with local behaviour
\be
  \Psi \propto (D- D_\star)^{\,\rho} \,,   
\ee
with $\rho$ being a non-integer singularity exponent (if $\rho$ is positive, then derivatives of $\Psi$ blow up). Indeed, as shown by \cite{DombSykes:1957,vanDyke1974}, 
for such singularities the ratio
\be
  \|\fett{\psi}^{(n)}\|/\|\fett{\psi}^{(n-1)}\| = \frac{1}{D_\star} \left[ 1- (1+\rho) \, \frac 1 n  \right]
\ee
is exactly linear for large $n$, which furthermore reveals the large-$n$ asymptotic behaviour $\| \fett{\psi}^{(n)} \|  \simeq \gamma \,n^{-\rho-1} \exp(\beta n)$, with $\beta$ and $\gamma$ being fitting coefficients \citep{PODVIGINA2016320}. See  Table~\ref{tab:extrapol} for the estimated values of $\rho$, and the supplementary material~\ref{sec:pod} for further details. We find that for most cases considered we have  $0 < \rho \lesssim 1$ (and rarely $\rho >1$), thereby indicating that the first time derivative of the displacement -- a.k.a.\ the velocity -- blows up at $D_\star$. We stress again that $D_\star$ is a critical time, which for $\Lambda$CDM  appears to be always later than the shell-crossing time, i.e., $D_\star > D_{\rm sc}$. Of course, cases when the velocity blows up at $D_\star$ are not physical and just reflect the strength of the underlying singular problem.

Strictly speaking, investigating the singularities in LPT should be done on a case-by-case basis for given random seed, as the outcome of such analysis can depend crucially on the given topology at hand. For example, we find that the singularity exponent is closer to zero if the local initial overdensity has quasi-spherical shape \citep[where LPT converges relatively slowly, cf.][]{Saga:2018,Rampf:2017tne}, while in that case $z_\star$  is just shortly behind $z_{\rm sc}$. By contrast, for initial overdensities that are rather close to being quasi-one-dimensional \citep[where LPT converges fast, see][]{Rampf:2017jan}, $\rho$ is larger while $z_\star \ll z_{\rm sc}$; see the supplementary material~\ref{app:outliers} for specific case examples. Nonetheless, the first shell-crossing  usually arises from rather quasi-spherical initial overdensities (see the next section).

Finally, let us comment on why in the CDM case we include ``only'' 15th-order effects, while we go up to 40th order for WDM. As mentioned above, the  $\| \fett{\psi}^{(n)}(\fett{q})\|$'s grow exponentially for large orders $n$ at shell-crossing locations, and thus lead to sharp peaks in Lagrangian space with exponentially increasing height at increasing perturbation order. Since the numerical computations of LPT coefficients involves Fourier transforms, those  peaks in $\| \fett{\psi}^{(n)}(\fett{q})\|$ can lead to wave-like features in real space due to the Gibbs phenomenon, thereby obscuring the LPT coefficients with unphysical artefacts at very large Taylor orders. These peaks are the sharpest for CDM spectra while they are tamer for WDM spectra, basically since the sharpness of those peaks is related to the information content of the initial density spectra at small spatial scales, where WDM spectra have less support. As a conclusion, we find that the Domb--Sykes plot for CDM  is unaffected by the Gibbs phenomenon only for $n \lesssim 15$ (see supplementary material~\ref{app:outliers} for details), while for WDM  we could not detect any such limitation. The connection between Fourier series convergence and LPT convergence in these cases could be an interesting aspect of future investigations.

\section{Which seeds/shapes collapse first?}\label{sec:collapsestructures}

As a last point we investigate which seeds shell-cross first. 
In that context it is 
useful to consider the Jacobian matrix $\fett{\rm J} \equiv \mathbb{1} + \nabL\fett{\psi}$ to a given perturbative truncation, and diagonalize its symmetric part into the coordinate system along the fundamental axes. The symmetrized Jacobian, denoted with $J^{\rm s}$, can then be written in terms of the three real eigenvalues $\lambda_{1,2,3}= \lambda_{1,2,3}(\fett{q},t)$ as
\be
 J^{\rm s} = 
  \left| 
    \begin{matrix}
       \,\lambda_1 \,& 0 & 0 \\
       0   & \, \lambda_2 \, &  0 \\
       0   & 0 & \, \lambda_3 \,    
    \end{matrix} 
 \right| \,,
\ee
where we impose the ordering $\lambda_{1}  \leq  \lambda_2  \leq \lambda_3$. Locally, shell-crossing is reached when  $\lambda_1 \to 0$, while instantaneous multi-lateral compressions with $\lambda_{1,2} \to 0$, or even $\lambda_{1,2,3} \to 0$, are essentially excluded for random initial conditions \citep{Zeldovich:1970,1970Ap......6..320D}.
Note that the above $\lambda_i$'s  relate to the eigenvalues   of the so-called deformation tensor $\nabL\fett{\psi}$, dubbed $\lambda_{i, {\rm d}}$, according to $\lambda_{i, \rm d} = \lambda_i -1$.

\begin{figure}
	\centering
	\includegraphics[width=0.95\columnwidth]{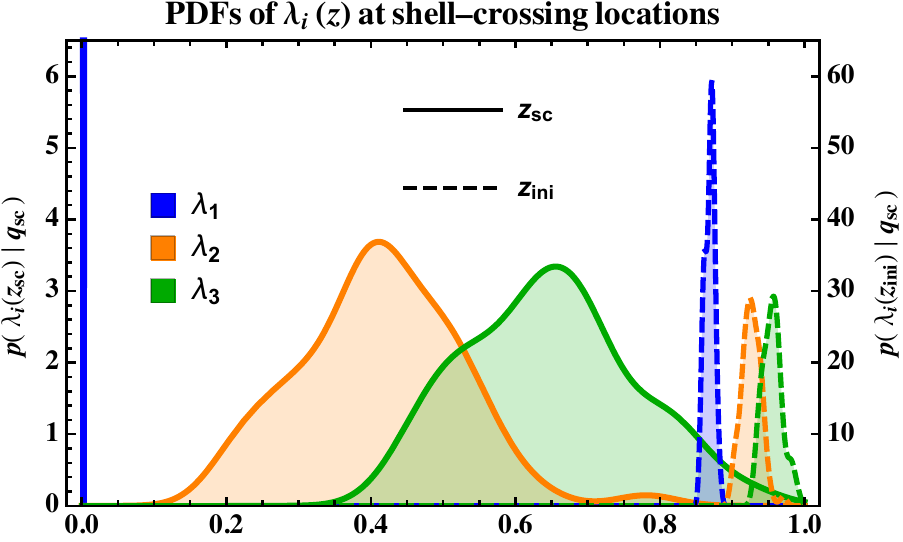}
	\caption{Conditional distribution of eigenvalues of $J$ from  locations $\fett{q}=\fett{q}_{\rm sc}$ that shell-cross first, obtained from 65 standard-$\Lambda$CDM realizations. Shown are the initial distributions at~$z_{\rm ini}=100$ (dashed lines; amplitudes scale according to right $y$-axis)  and at shell-crossing~$z_{\rm sc}$ (solid lines; left $y$-axis). }
	\label{fig:PDFS}
\end{figure}

In Fig.\,\ref{fig:PDFS} we show the resulting distribution of $\lambda_i$'s at the initial redshift $z_{\rm ini} = 100$  (dashed lines) and shell-crossing redshift $z_{\rm sc}$ (solid lines), obtained for 65 standard $\Lambda$CDM random realizations of first shell-crossing locations. To be specific, for a given realization we search for the first shell-crossing location and evaluate the eigenvalues at those Lagrangian positions at redshift  $z_{\rm sc}$ to 5th-order precision, and to second order at $z_{\rm ini}=100$. We remark that $z_{\rm sc}$ varies slightly, depending on the given seed, while we keep $z_{\rm ini}$ fixed. 

A prominent feature in Fig.\,\ref{fig:PDFS} is the distribution of $\lambda_1$ at $z_{\rm sc}$, which is, as expected,  essentially a delta peak at the origin. The other two eigenvalues have  medians of $0.41$
and $0.65$ respectively, indicating that the first collapsing objects  have fairly ellipsoidal shape initially. Another noticeable but subtle feature is the appearance of a  smaller peak in $\lambda_2$ around $0.8$ at $z_{\rm sc}$ (that actually traces back to another small peak in $\lambda_2 \simeq 0.96$ at $z_{\rm ini}$), which implies that quasi one-dimensional collapse rarely is the source of the first shell-crossing.
The initial distributions of the $\lambda_i(\fett{q}_{\rm sc}, z_{\rm ini})$'s turn out to be already fairly extreme, with at least $3\sigma$ deviations of $\lambda_{2,3}(\fett{q}_{\rm sc}, z_{\rm ini})$  compared to $\lambda_{2,3}(\fett{q}, z_{\rm ini})$ from all grid points,  and  at least $5\sigma$ for $\lambda_1$.

Finally, we have also investigated the ellipticity $e = (\lambda_{1, \rm d} - \lambda_{3, \rm d})/2\sum \lambda_{i,\rm d}$ and prolateness $p= (\lambda_{1, \rm d} - 2 \lambda_{2, \rm d} + \lambda_{3,\rm d})/2\sum\lambda_{i,\rm d}$ of the collapsing seeds in full non-linearity (see  supplementary material~\ref{app:eandp} for linear predictions). For the first shell-crossing locations we find that both $e$ and $p$ change hardly between $z_{\rm ini}$ and $z_{\rm sc}$, but with a slight tendency in becoming less oblate/prolate at late times. This tendency is also reflected by a slight shift of their medians of  $e = 0.171$ and $p=0.054$ at~$z_{\rm ini}$, and $e = 0.168$ and $p=0.042$ at~$z_{\rm sc}$. Further details are provided in the supplementary material~\ref{app:eandp}.

We remark that the present avenue is fairly distinct from the ones of e.g.\ \cite{1970Ap......6..320D,Bardeen1986,Sheth2001}: While those works considered the statistics of peaks by means of linear theory or approximative collapse models, here we investigate the matter collapse on a deterministic and fully non-linear level for various seeds, from which we are able to deduct what the (extreme) statistics of those shell-crossing regions are.

\section{Concluding remarks}

We have shown that Lagrangian perturbation theory can be used to rigorously investigate the first shell-crossing in a $\Lambda$CDM Universe. 
We find that the LPT series for the displacement field converges until shell-crossing and, accidentally, even shortly after. Actually, the latter statement guarantees a rather fast convergence until shell-crossing, thereby allowing to determine time and location of the first shell-crossing efficiently and to high accuracy. 

The precise time of first shell-crossing varies with chosen numerical resolution and spectrum of the initial data. In fact, 
fixing the mesh spacing fixes also an effective fluid description. Below the grid scale this effective fluid is blind to large-$k$ fluctuations in the spectrum of primordial density fluctuations.
Whether such so-called UV effects should be included in cosmic fluid descriptions (and in $N$-body simulations)  is subject to debate \citep[e.g.,][]{Taruya:2018,2020arXiv200211357C} and should be assessed in forthcoming work.

Our results can be applied to generate very accurate initial conditions for numerical $N$-body simulations,  possibly even for simulations with multiple fluids \citep[cf.][]{Hahn:2020,Rampf:2020}. Higher-order initial conditions are not only more accurate but also open up the portal for late-time initializations, thereby suppressing strongly discretization errors that affect such simulations at earlier times \citep[see e.g.,][]{Michaux:2020}. 
In the longterm, our methods make it possible to close the gap between theoretical and numerical methods for the cosmic large-scale structure. Possible future directions include the study of LPT against numerical simulation techniques, as well as investigating the fully non-linear formation process of primordial dark-matter halos within the tidal gravitational field.

\section*{Acknowledgements}

We thank Thomas Buchert, Uriel Frisch and Patrick McDonald for useful discussions and/or comments on the manuscript.
C.R.\ is a Marie Sk\l odowska-Curie Fellow and acknowledges funding from the People Programme 
of the European Union Horizon 2020 Programme under Grant Agreement No.\ 795707 (COSMO-BLOW-UP). 
 O.H.\ acknowledges funding from the European Research Council 
(ERC) under the European Union's Horizon 2020 research and innovation programme, 
Grant agreement No. 679145 (COSMO-SIMS).

\section*{Data Availability}

The software implementing the recursion relations is freely available at \url{https://bitbucket.org/ohahn/monofonic}.
The data underlying this article will be shared on reasonable request to the corresponding author.



\bibliographystyle{mnras}
\bibliography{bibliography} 



\appendix

\section*{{\it Supplementary material}}

\vskip0.3cm

\section{Numerical implementation}\label{app:implementation}

\subsection{Employed algorithms}

The recursion relations can be broken down into a loop over kernels operating on vector fields. Two of these kernels $\mu_2(\cdot,\cdot)$ and $\mu_3(\cdot,\cdot,\cdot)$ are scalar valued and contribute to the longitudinal displacement (Eq.\,\ref{eq:scalarrec}). Pseudo-code algorithms to compute them are given in Algorithms~\ref{alg:mu2} and \ref{alg:mu3}, respectively. \footnote{We remark that algorithm~\ref{alg:mu3} only delivers $\mu_3^{(n_1,n_2,n_3)}$ when summed implicitly according to $\sum_{n=n_1+n_2+n_3}$.}

A third vector-valued kernel, $\fett{C}(\cdot,\cdot,\cdot)$, contributes  to the transversal displacement (needed for Eq.\,\ref{eq:cauchy}); its algorithm is given in Algorithm~\ref{alg:C}. Note that these kernels operate in Fourier space on fields which are functions of wavevectors, $\fett{A}=\fett{A}(\fett{k})$. The operators are also in Fourier space so that $\partial_j := {\rm i}\,k_j$, and $\nabla^{-2} := -\|\fett{k}\|^{-2}$ if $\|\fett{k}\|\neq0$ and zero otherwise. All products between terms are therefore indeed convolutions, denoted by the star symbol `$\star$'. 

The final algorithm, looping over all orders $i$ up to the desired truncation order $n$ is given in Algorithm~\ref{alg:nlptrecursion}, where the recursion has been algorithmically realised as a loop.\footnote{In line 10  of algorithm~\ref{alg:nlptrecursion}, $\fett{T}$ does not receive any contribution since the transverse field vanishes when $j= i-j$ (i.e., since in that case, for any integer $X$, we have $\varepsilon_{ijk} \psi_{l,j}^{(X)} \psi_{l,k}^{(X)}=0$).
 Furthermore, in line 13, the double sum arises from the identity $\sum_ {k + l + m =  i} a_ {klm} = \sum_ {k = 1}^{i - 2}\sum_ {l = 1}^{i - 1 - k} a_ {kl (i - k - l)}$.}
Once the displacement is determined to the desired order, the truncated Jacobian can be determined using Algorithm~\ref{alg:J}.

\begin{algorithm}
\caption{\label{alg:mu2}Calculate $\mu_2(\fett{A},\fett{B})$}
\begin{algorithmic}[1]
\Require vector fields $\fett{A}=\{A_{0\dots2}\}$ and $\fett{B}=\{B_{0\dots2}\}$
\State $\mu_2 \gets 0$
\If{$\fett{A}=\fett{B}$}
 \State $\mu_2 \gets \partial_0 A_0 \star ( \partial_1 A_1  + \partial_2 A_2 )
    + \partial_1 A_1 \star \partial_2 A_2 - \partial_1 A_0 \star \partial_0 A_1
    - \partial_2 A_0 \star \partial_0 A_2 - \partial_2 A_1 \star \partial_1 A_2 $
\Else
\For{$i$}{0}{2}
\State $j \gets (i+1) \mod 3$
\State $k \gets (i+2) \mod 3$
\State $\mu_2 \gets \mu_2 + \partial_i A_i \star ( \partial_j B_j +  \partial_k B_k )  -  \partial_i A_j \star \partial_j B_i  -  \partial_i A_k \star \partial_k B_i$
\EndFor  
\State $\mu_2 \gets \mu_2/2$
\EndIf
\end{algorithmic}
\end{algorithm}

\begin{algorithm}
\caption{\label{alg:mu3}Calculate $\mu_3(\fett{A},\fett{B},\fett{C})$}
\begin{algorithmic}[1]
\Require vector fields $\fett{A}=\{A_{0\dots2}\}$, $\fett{B}=\{B_{0\dots2}\}$ and $\fett{C}=\{C_{0\dots2}\}$
\State $\mu_3 \gets 0$
\For{$i$}{0}{2} 
\State $j \gets (i+1) \mod 3$
\State $k \gets (i+2) \mod 3$
\State $\mu_3 \gets \mu_3 + \partial_i A_0 \star  \left( \partial_{j} B_{1} \star \partial_{k} C_{2} - \partial_{k} B_{1} \star \partial_{j} C_{2}\right)$ 
\EndFor
\end{algorithmic}
\end{algorithm}

\begin{algorithm}
\caption{\label{alg:C}Calculate  $\fett{C} = \nabla A_l \times \nabla B_l$ }
\begin{algorithmic}[1]
 \Require vector fields $\fett{A}=\{A_{0\dots2}\}$ and $\fett{B}=\{B_{0\dots2}\}$
 \State $\fett{C} \gets \fett{0}$
 \For{$i$}{0}{2} 
  \State $j \gets (i+1) \mod 3$
  \State $k \gets (i+2) \mod 3$
  \For{l}{0}{2}
    \State $C_i \gets C_i +  \partial_j A_l \star \partial_k B_l -  \partial_k A_l \star \partial_j B_l$
  \EndFor
 \EndFor
\end{algorithmic}
\end{algorithm}

\begin{algorithm}
\caption{\label{alg:nlptrecursion}Calculates all $n$LPT displacements $\fett{\Psi}^{(i)}$ for $i \leq n$ from initial gravitational potential $\varphi^\ini$}
\begin{algorithmic}[1]
\Require scalar field $\varphi^\ini$, $n\ge1$, temporary storage for $L$ and $\fett{T}$
\State $\fett{\Psi}^{(1)} \gets -\fett{\nabla} \varphi^\ini $
\For{$i$}{2}{$n$} \Comment{Loop over order $n$}
\State $L\gets0 $   \Comment{longitudinal displacement}
\State $\fett{T}\gets \fett{0} $    \Comment{transverse displacement}
  \For{$j$}{1}{$i/2$}
   \If{$j \neq i-j$} \Comment{uses symmetries}
     \State $L\gets L+2\,\frac{(3-i)/2-j^2-(i-j)^2}{(i+3/2)(i-1)} \,\mu_2( \fett{\Psi}^{(j)},\, \fett{\Psi}^{(i-j)})$ 
     \State $\fett{T} \gets  \fett{T} + \left( 1 - 2j/i \right) \,\fett{C}(\fett{\Psi}^{(j)},  \fett{\Psi}^{(i-j)})$
   \Else   
     \State $L\gets L+\frac{(3-i)/2-j^2-(i-j)^2}{(i+3/2)(i-1)} \,\mu_2( \fett{\Psi}^{(j)},\, \fett{\Psi}^{(i-j)})$ 
   \EndIf
  \EndFor
    \For{$k$}{1}{$i-2$} 
     \For{$l$}{1}{$i-1-k$}
        \State $L\gets L +  \frac{(3 - i)/2 - k^2 - l^2 - (i - k - l)^2}{ (i+3/2)( i-1)} \,\mu_3(\fett{\Psi}^{(k)}, \fett{\Psi}^{(l)},\fett{\Psi}^{(i-k-l)}) $
     \EndFor
  \EndFor
  
  \For{$d$}{0}{$2$} \Comment{Loop over dimensions}
    \State $e \,\gets (d+1) \mod 3$
    \State $f \gets (d+2) \mod 3$
    \State $\Psi_d^{(i)} \gets \nabla^{-2} \left( \partial_d L  - ( \partial_e T_f - \partial_f T_e) \right) $ \Comment{Helmholtz}
  \EndFor 
\EndFor \Comment{$\fett{\Psi} = \sum_{i=1}^n \fett{\Psi}^{(i)}  D_+^i$}
\end{algorithmic}
\end{algorithm}

\begin{algorithm}
\caption{\label{alg:J}Calculate truncated Jacobian $J = \det \fett{\rm J}$ to order $n$}
\begin{algorithmic}[1]
 \Require $\fett{\Psi}^{(n)}$
 \State ${\fett{\rm J}}, \fett{\Psi} \gets \fett{0}$
 \For{$s$}{0}{$n$} 
   \State $\fett{\Psi} \gets  \fett{\Psi} +  \fett{\Psi}^{(s+1)} \, D^{s+1}$
 \EndFor
 \State $\fett{\rm J} \gets \nabL \fett{\Psi}$
 \State $J_{00} \gets J_{00} +1$
 \State $J_{11} \gets J_{11} +1$
 \State $J_{22} \gets J_{22} +1$
 \State $J_{00} \gets  \det \fett{\rm J}$ \Comment{returns $J$}
\end{algorithmic}
\end{algorithm}

\subsection{Corner modes and Gibbs phenomenon}\label{app:corner}

In numerical cosmology it is customary to obtain a realization of Gaussian fields on grid points by first convolving the input transfer functions $T(k)$ (or power spectra $P(k)$) in real space with a Gaussian white noise. While the (one-dimensional) input $T(k)$ or $P(k)$ is isotropic in Fourier space, its discretized version on Cartesian grid points is not, since those modes are only set for $0 < (k_x,k_y,k_z) < k_{\rm Ny}$ where $k_{\rm Ny} = \uppi N^{1/3}/L_{\rm box}$ is the particle Nyquist frequency. This implies that non-zero modes $k > k_{\rm Ny}$ are naturally excited in this process, which are usually called corner modes; see \cite{2017ApJ...837..181F} for an extensive discussion.

We find that higher-order $n$LPT terms ($n\geq 6$)  suffer from the Gibbs phenomenon if corner modes are not filtered out. This is likely due to missing counter modes for these modes and possibly can be proven rigorously. There seems to be no problem in Fourier space, no odd features appear in the spectra, but the phenomenon appears in real space. It completely vanishes if we apply a sharp $k$-space top hat filter in a very final post-processing step in the following way:
\be
   \widehat{\boldsymbol{\Psi}}(\boldsymbol{k})  \to \widehat{\boldsymbol{\Psi}}(\boldsymbol{k})  \times 
 \left\{ 
    \begin{array}{ll}
      1 & \textrm{if}\quad \|\boldsymbol{k}\| < k_{\rm Ny} \\
      0 & \textrm{otherwise}
    \end{array} 
  \right. \,.
\ee

\begin{figure}
	\centering
	\includegraphics[width=0.99\columnwidth]{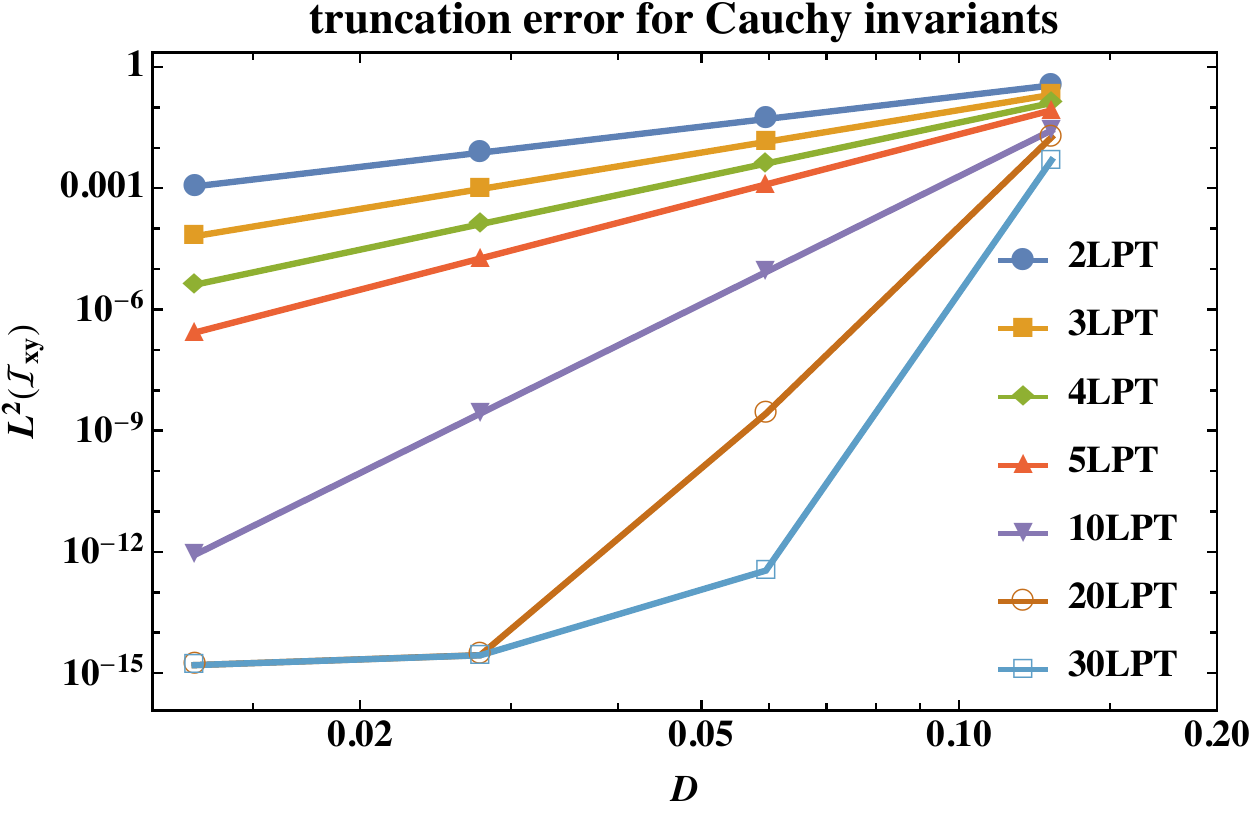}
	\caption{RMS of spurious vorticity for random $\Lambda$CDM initial conditions: $L^2$ norm of the invariant $\mathcal{I}_{xy}$ as a function of the linear theory growth function $D$. For a correct implementation of $n$LPT, one expects the $D^n$ scaling as seen here. The 20LPT and 30LPT curves level off at machine precision. }
	\label{fig:invariant_scaling}
\end{figure}

\subsection{Numerical tests and convergence studies}\label{sec:num-and-theo-convergence}

Numerical implementations of LPT with $\Lambda$CDM random initial conditions are 
a challenge, that so far have been only addressed to third order \citep[see e.g.][]{Michaux:2020}. In the following we provide more details on the employed numerical tests.

\paragraph*{Perturbative conservation of Cauchy invariants.} The curl equation~\eqref{eq:cauchy} appearing in the LPT recursion relations exploits essentially the conservation of Eulerian vorticity $\nabla_x \times \fett{v}$, which formulated in Lagrangian space turn into the so-called Cauchy invariants $C_p$
\be \label{eq:cauchyapp}
   C_p = \varepsilon_{pln} \dot x_{k,l} x_{k,n} = 0 \,,
\ee
where the ``0'' is a nonperturbative statement. In perturbative refinements, however, it is clear that the vanishing of the Cauchy invariants can only guaranteed in a perturbative sense.
Indeed, as \cite{Uhlemann:2019} have shown, at fixed order $n>1$, LPT excites artificial vortical modes at order $n+1$.
The expected truncation error at order $n$ in LPT scales with powers of $D_+^n$ (since Eq.\,\eqref{eq:cauchyapp} contains a time derivative). By contrast, incorrect numerical implementations of LPT (e.g., not de-aliased) can lead to different scalings \citep{Michaux:2020}.

In Fig.\,\ref{fig:invariant_scaling} we show the resulting scaling for our implementation, by means of the $xy$ component of the antisymmetric tensor ${\cal I}_{ij} := \varepsilon_{pij} C_p = \dot x_{k,i} x_{k,j} - \dot x_{k,j} x_{k,i}$, which indeed show the correct scaling behaviour.

\paragraph*{Numerical convergence/resolution studies.} Next we investigate the numerical convergence of our results by varying box size and/or resolution; for the mathematical convergence of LPT, which is a distinct problem involving subsequent higher LPT orders, see the  following paragraph.  Henceforth we evaluate certain fields and conditions only at the truncated order $n=3$.

\begin{figure}
 \begin{subfigure}{.45\textwidth}
	\centering
	\includegraphics[width=0.97\columnwidth]{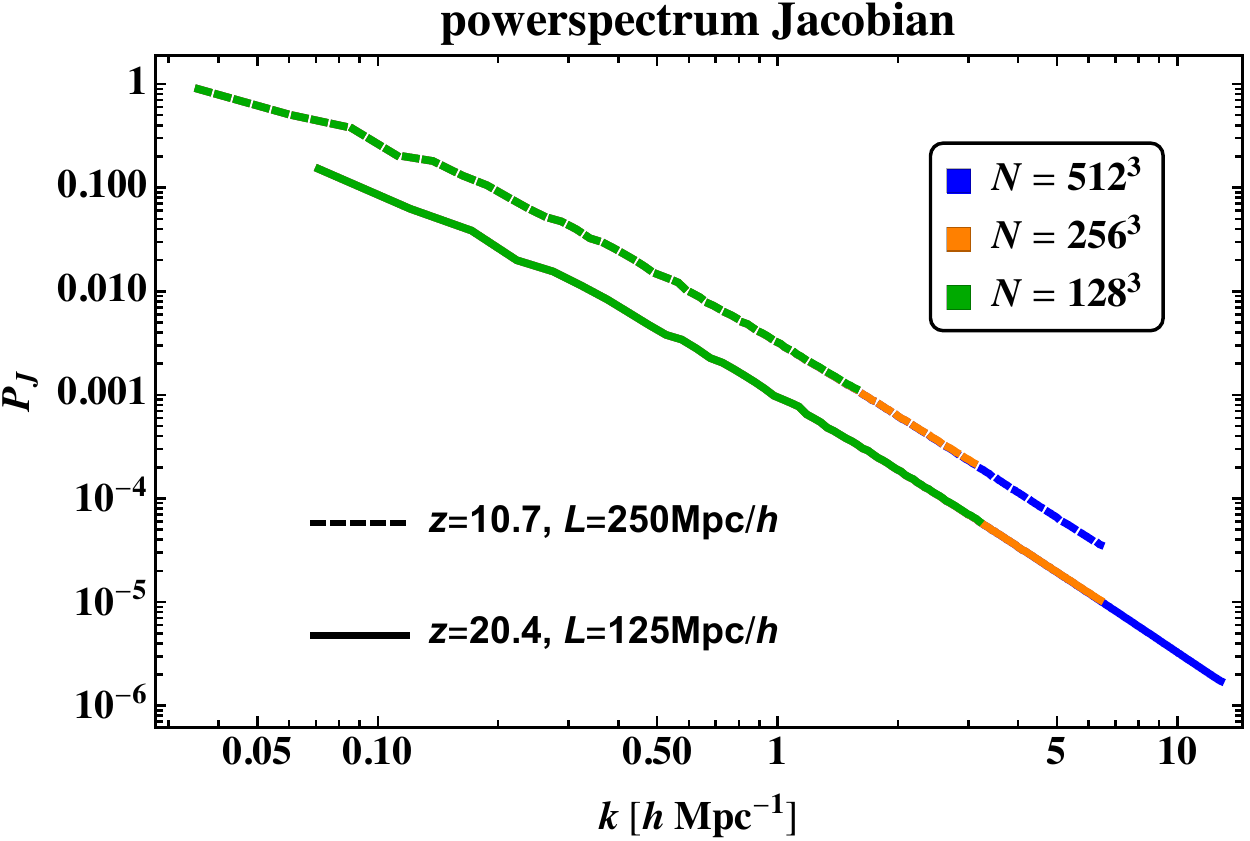}
 \end{subfigure}
 \begin{subfigure}{.45\textwidth}
   \vskip-0.5cm\hskip0.15cm	\includegraphics[width=0.98\columnwidth]{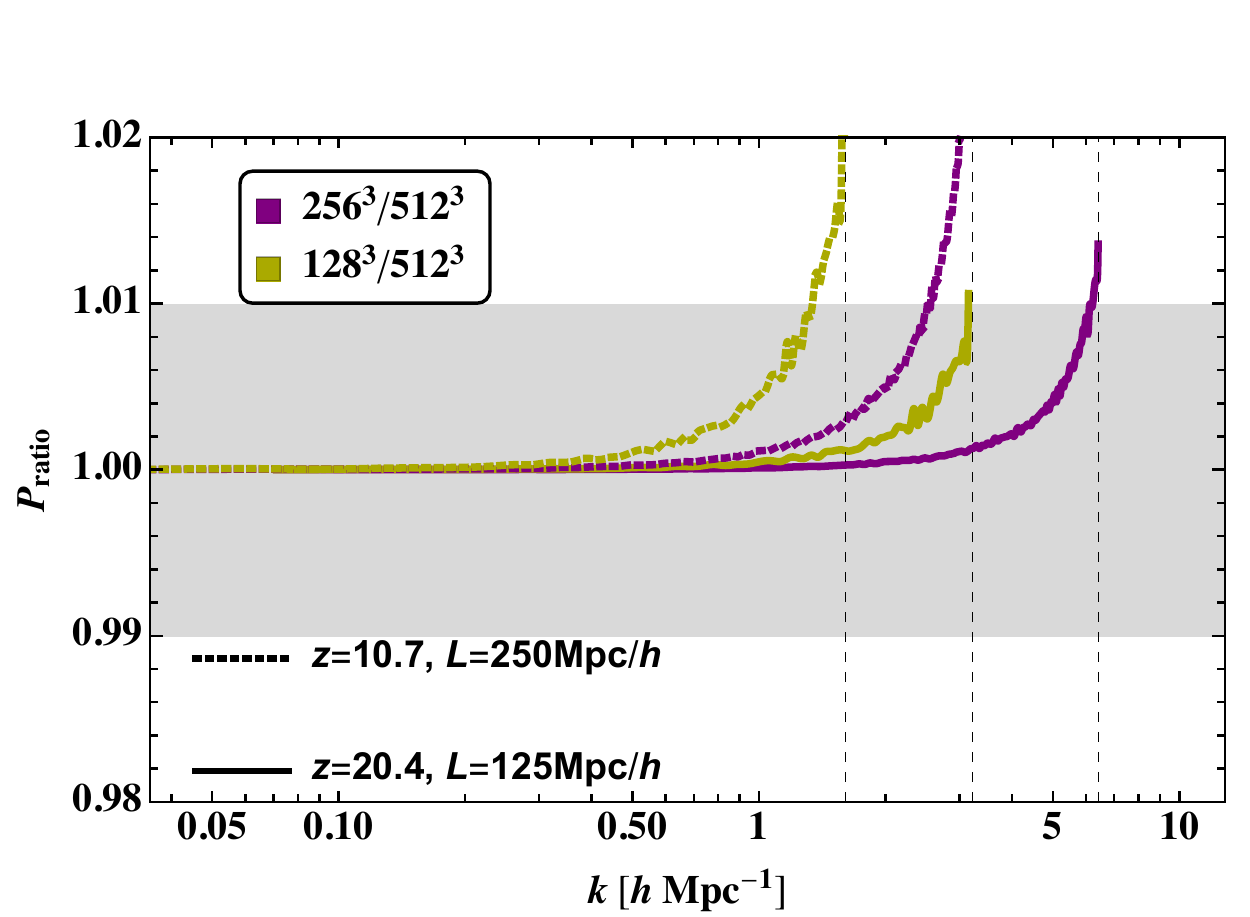}
 \end{subfigure}
 \caption{{\bf Top panel:} Power spectrum $P_J$ of Jacobian at various resolution settings, evaluated at the respective redshift when the run with the highest resolution shell-crosses first. Note that the shell-crossing time for CDM can differ  (see also Figs.\,\ref{fig:zscdependendence}), nonetheless these results demonstrate that the LPT predictions for observables are unaffected by this mismatch. {\bf Bottom panel:} Ratios of power spectra for resolutions as above. The vertical dashed lines denote the relevant particle Nyquist frequencies.}
	\label{fig:PJ}
\end{figure}

Figure\,\ref{fig:PJ} shows the power spectrum of the Jacobian at 3LPT, for the resolutions
$N=128^3,\,256^3,\,512^3$ (respectively colored green, orange and blue) and for the box sizes $L_{\rm box} = 250\Mpch,\, 125\Mpch$ (respectively dashed and solid lines).
For given box size we evaluate the power spectra at the instance of the first shell-crossing for the highest resolution. We are doing so since we want to rule out that our numerical convergence studies are being hampered by the potential loss of LPT convergence.
As can be seen from the figure, numerical predictions are fully converged.

\begin{figure}
\centering
 \includegraphics[width=.82\columnwidth]{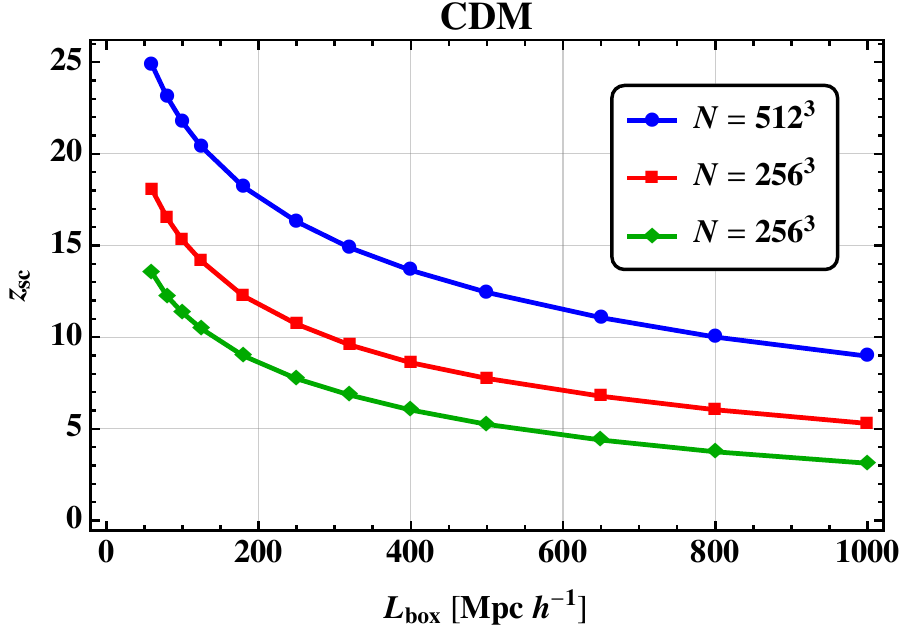}\\[0.5cm]

 \includegraphics[width=.8\columnwidth]{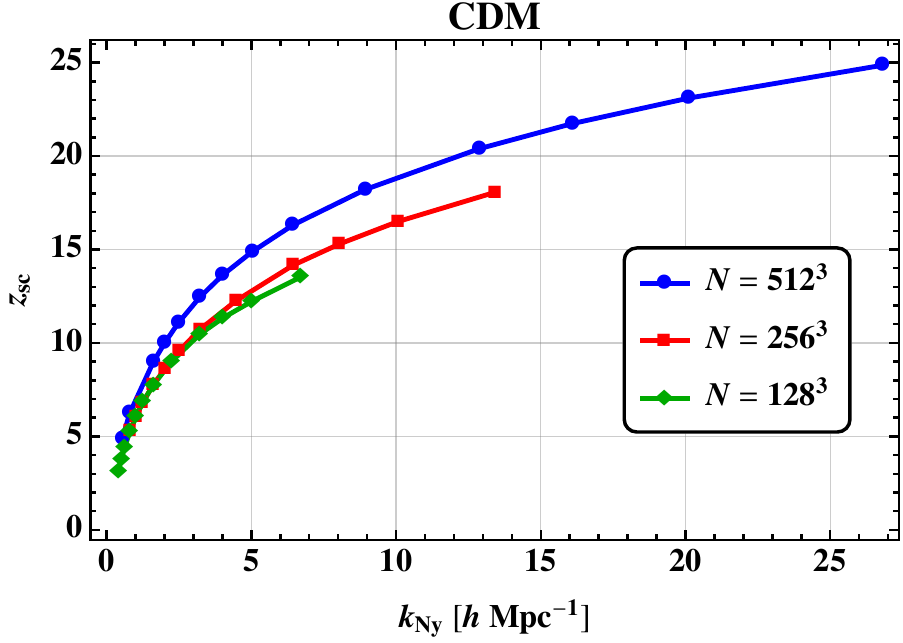}\\[0.5cm]

\includegraphics[width=.82\columnwidth]{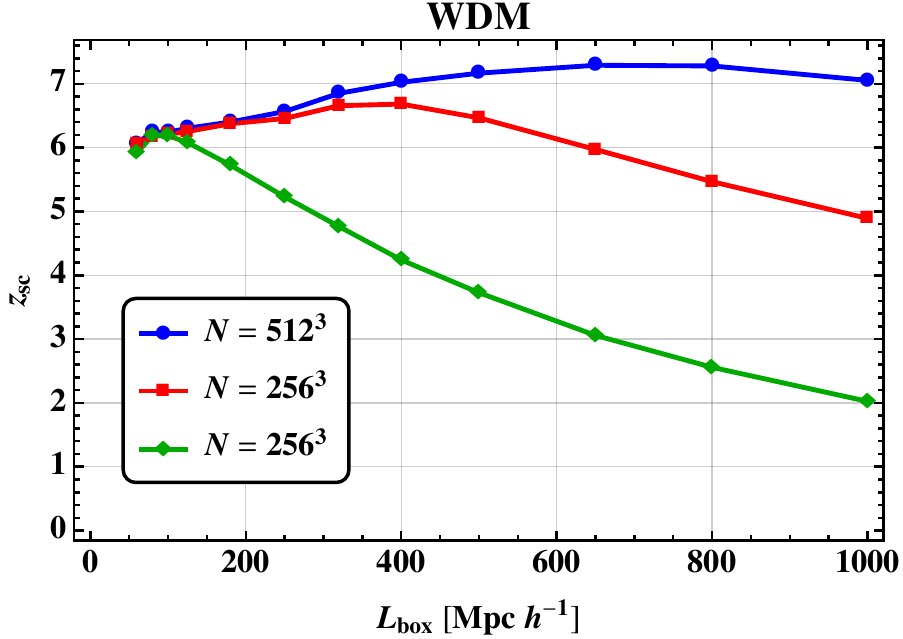}\\[0.5cm]
 
\includegraphics[width=.8\columnwidth]{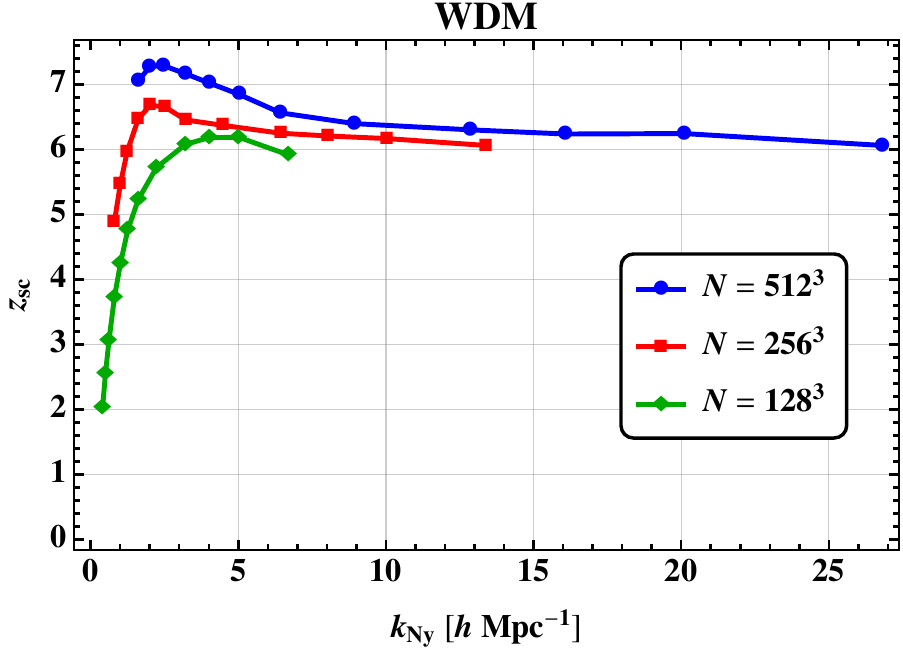}

\caption{Resolution dependence on the redshift of first shell-crossing $z_{\rm sc}$ for various resolution settings and cosmologies. {\bf Top panel:} Shift of $z_{\rm sc}$ as a function of box size $L_{\rm box}$ for $N=512^3, 256^3, 128^3$ number of grid points (from top to bottom, colored blue, green and orange) for CDM. {\bf Second panel:} Same as first panel  but as a function of $k_{\rm Ny}$. {\bf Third and fourth panels:} Same as the first panels but for WDM.}
\label{fig:zscdependendence}
\end{figure}

Next we investigate the resolution dependence on the shell-crossing time. As discussed in the main text, for a $\Lambda$CDM-like spectra,  setting the number of grid points and box size amounts to setting an effective fluid description: beyond the grid scale, LPT is effectively blind to more fine-grained details of the initial spectrum at large Fourier modes. Furthermore, the colder the dark matter is, the less compact the resulting spectrum is and, as a consequence, one expects a more pronounced dependence on the shell-crossing time for CDM than WDM. 
In Fig.\,\ref{fig:zscdependendence} we show this expected behaviour for various settings, either as a function of the box size $L_{\rm box}$ (left panels) or of the particle Nyquist frequency $k_{\rm Ny} = \uppi N^{1/3}/L_{\rm box}$ (right panels).

\begin{figure}
	\centering
	\includegraphics[width=0.85\columnwidth]{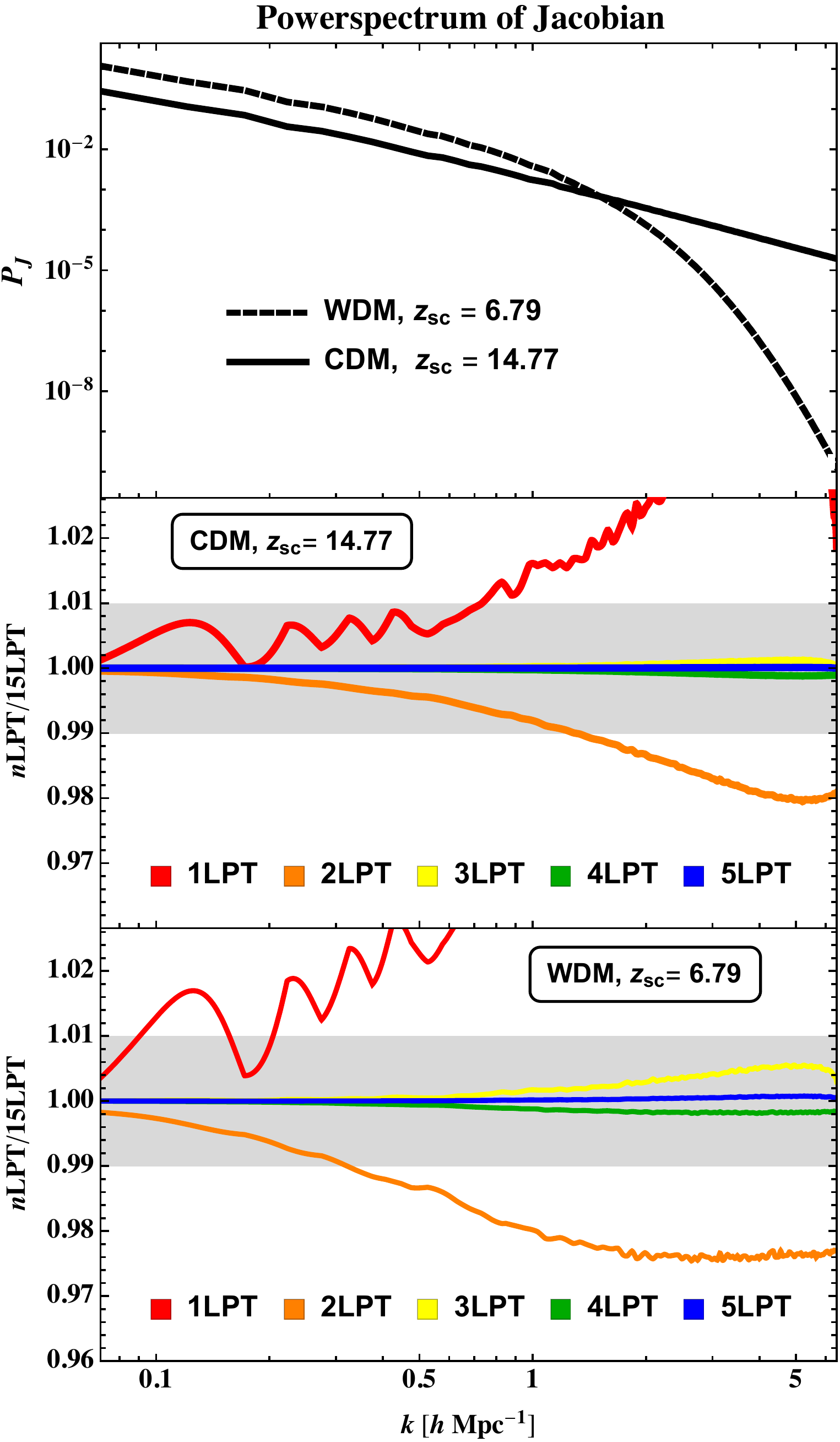}
	\caption{{\bf Top panel:} Power spectra $P_J$ of Jacobian $J= \det( \mathbb{1} + \nabL\fett{\psi})$ to 15th-order precision in LPT at the redshift $z_{\rm sc}$ of first shell-crossing where $J(z_{\rm sc})=0$, for CDM (solid black lines) and WDM with $m_{\rm wdm}=250\,$eV (dashed black lines). {\bf Middle panel:}  Ratios of power spectra $P_J$ at the first shell-crossing, to $n$th order versus 15LPT for CDM. {\bf Lower panel:} Same as middle panel but for WDM. For both CDM and WDM, convergence to sub per cent precision (denoted by grey shading) is achieved for $n \geq 3$. }
	\label{fig:powerJ}
\end{figure}

\paragraph*{Theoretical convergence, displacement norms and beyond the first shell-crossing.} Here we show explicit results related to the LPT convergence of the Jacobian at $z=z_{\rm sc}$, 
and provide figures that show the generic trends of the displacement norms (relevant for generating the Domb--Sykes plot).

In Fig.\,\ref{fig:diffJ} we have demonstrated the convergent behaviour of LPT at shell-crossing by means of real-space statistics. 
This convergence can also be observed in the power spectrum of the Jacobian. The top panel in Fig.\,\ref{fig:powerJ} shows $P_J$ at order 15LPT for CDM  and WDM (respectively solid and dashed lines) for a given random realization, both evaluated at their converged redshifts $z_{\rm sc}$ of first shell-crossing (here $k_{\rm Ny} = 6.43 h\, \Mpc^{-1}$). In the middle (CDM) and lower (WDM) panels we show ratios of $P_J$ at order $n$LPT versus 15LPT, for $n=1,2,\ldots,5$; here it becomes evident that  sub per cent convergence of $P_J$ at shell-crossing is reached already at 3LPT.

\begin{figure}
	\centering
	\hskip-0.04cm\includegraphics[width=\columnwidth]{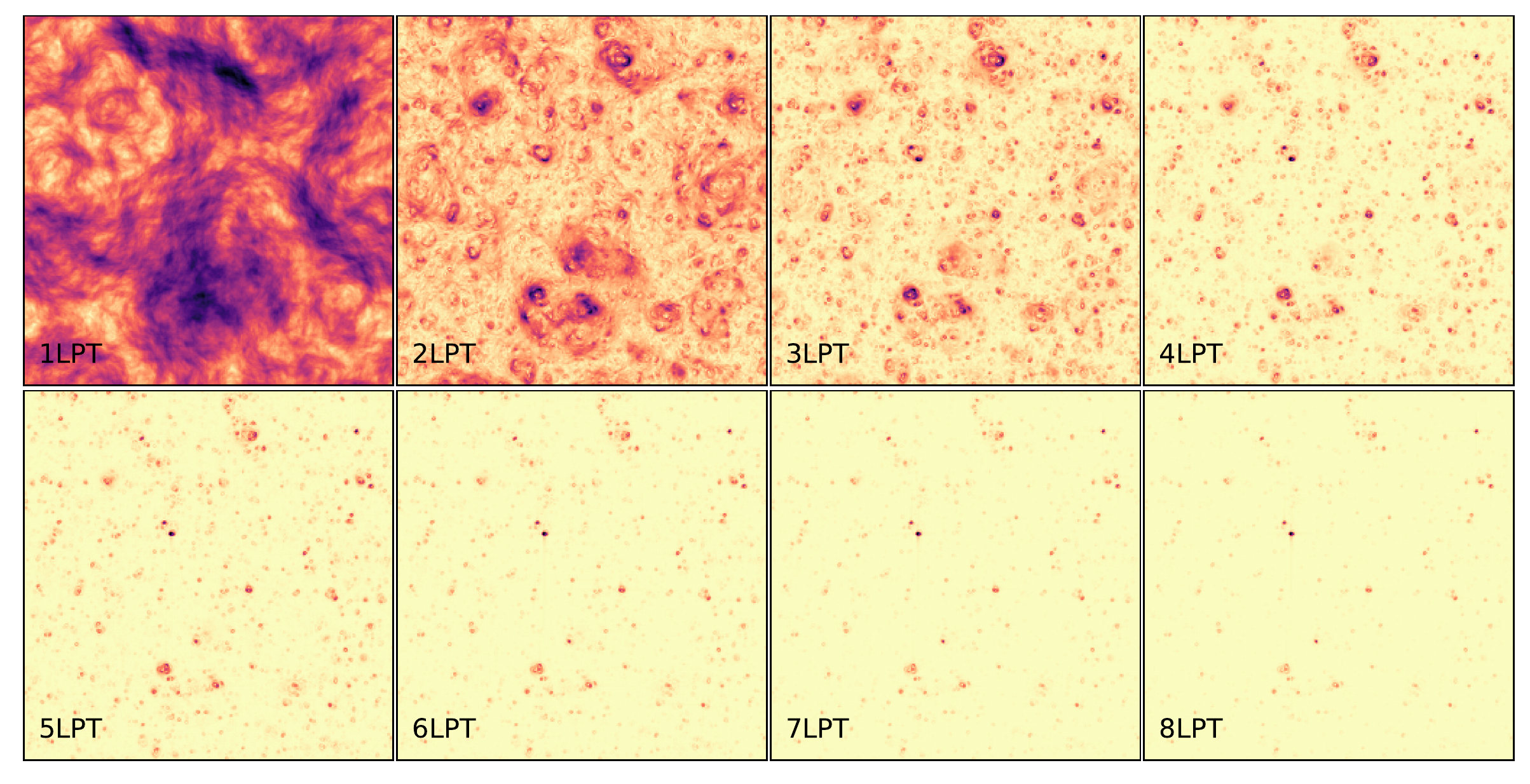}\\
	\includegraphics[width=0.985\columnwidth]{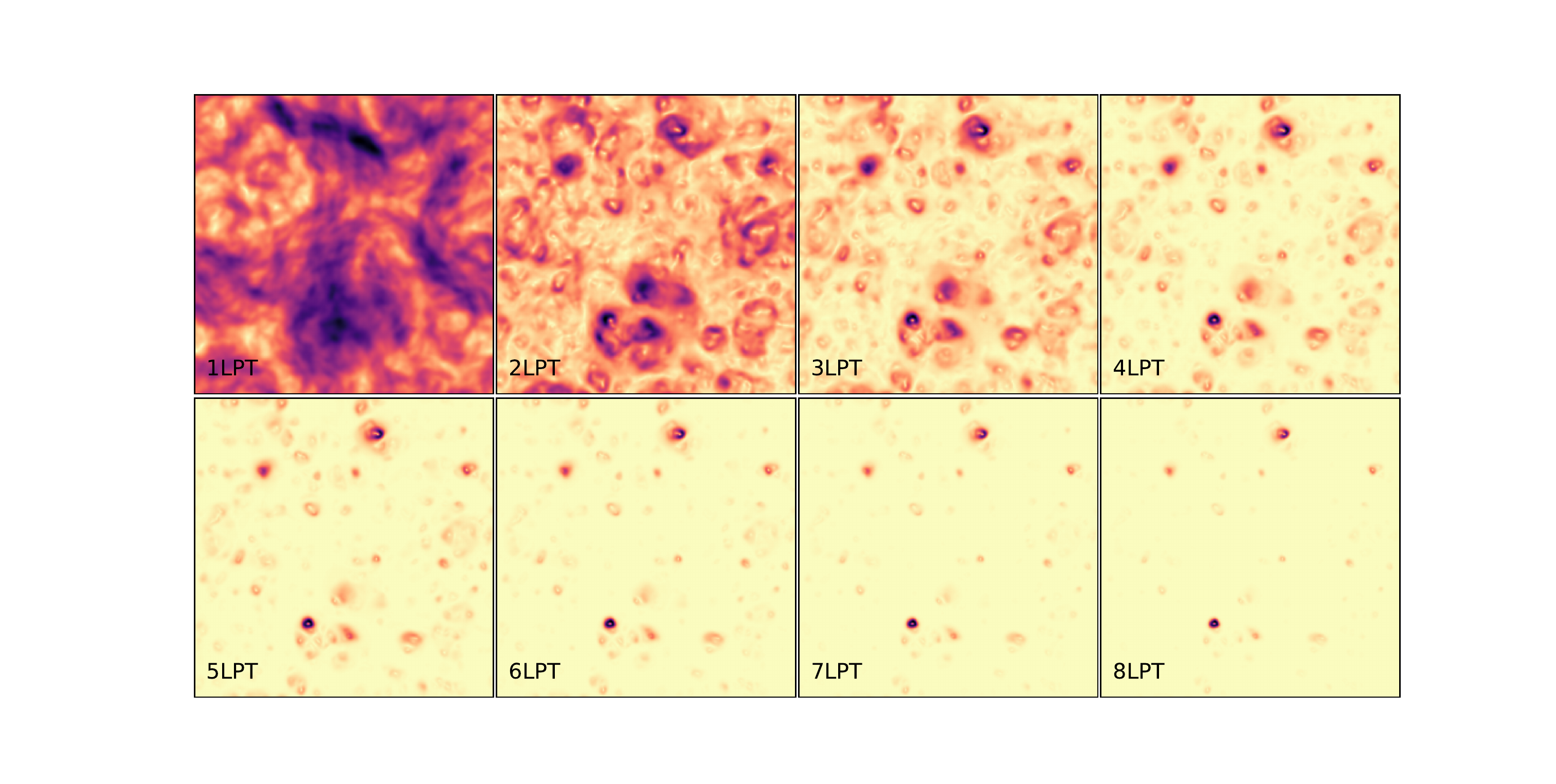}
	\caption{Norm of the displacement field $\| \boldsymbol{\psi}^{(n)} \|$ for the first eight orders of $n$LPT. Shown is a slice through a box of $128\,h^{-1}{\rm Mpc}$ width and height, calculated at a resolution of $256^3$. CDM top, WDM bottom.}
	\label{fig:displacementnorm}
\end{figure}

Finally, for reasons of completeness, we show in Fig.\,\ref{fig:displacementnorm} the norms of the displacement field  $\| \boldsymbol{\psi}^{(n)} \|$ for the first eight orders of $n$LPT, both for CDM and WDM.

\paragraph*{Loss of mathematical LPT convergence at late times.} In the main text we have determined the redshifts $z_\star$ for which the LPT series converges mathematically. Going just shortly after $z_\star$, the signature of losing convergence is rather subtle. Here, for reasons of completeness we show some LPT results at late times for which convergence is lost (and not even a close call).

\begin{figure}
	\centering
	\includegraphics[width=0.4\textwidth]{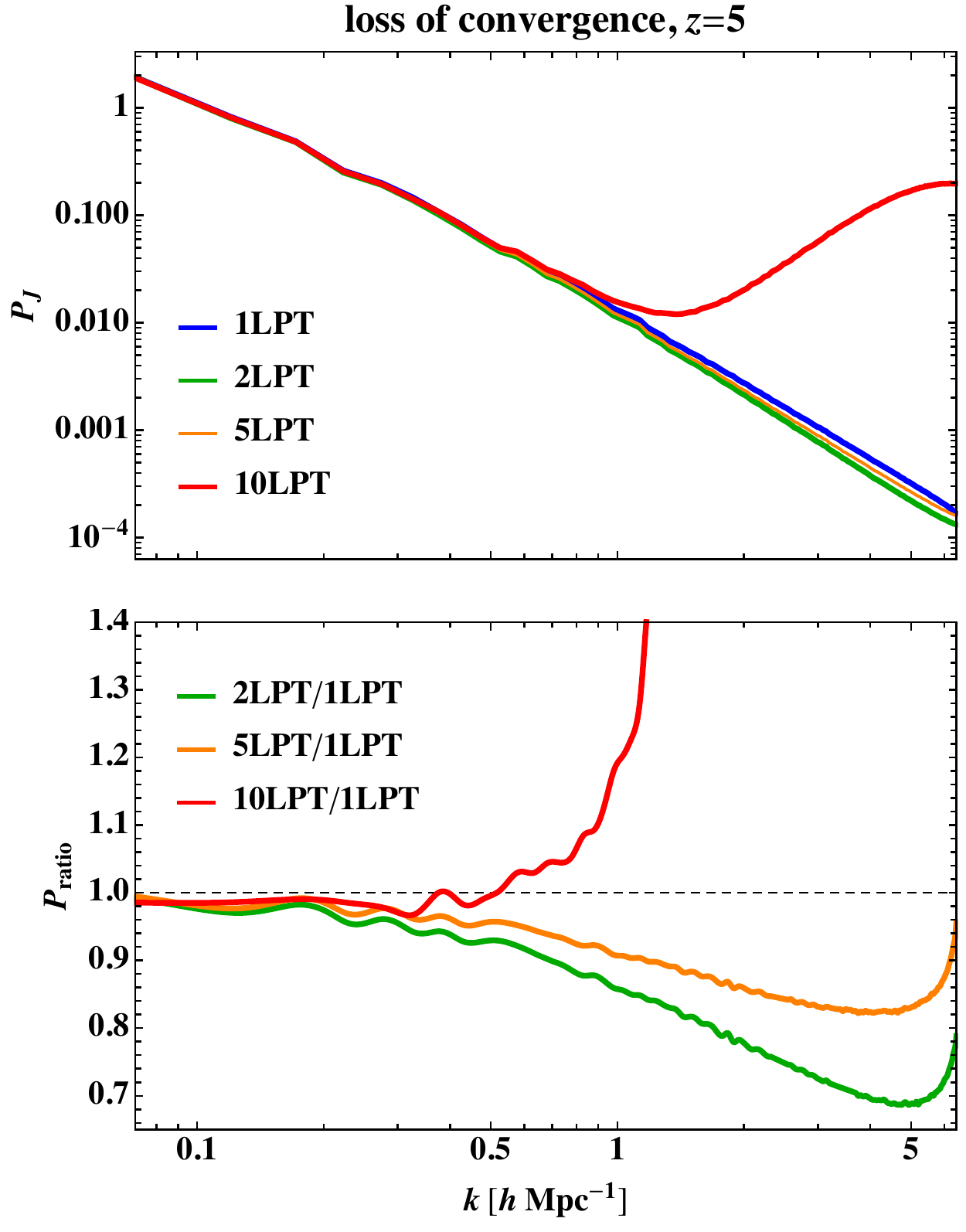}
        \caption{Power spectrum of Jacobian at various LPT orders (top panel), together with the ratios (lower panel). All power spectra are for CDM and evaluated at $z=5$, which is well beyond the estimated radius of convergence of $z_\star =10$. The particle Nyquist wave number is $k_{\rm Ny} = 6.43 h\, \Mpc^{-1}$. }
	\label{fig:PJdivlpt}
\end{figure}

In Fig.\,\ref{fig:PJdivlpt} we show the power spectrum of $J$ at $z=5$, for a grid of $N=256^3$  with box length $L_{\rm box} = 125 \Mpch$ for which shell-crossing occurs at $z_{\rm sc} = 15$ and convergence is lost at $z_\star=10$. Thus, results $z\leq 10$, as the ones displayed in the figure, are beyond the radius of convergence, where LPT displays divergent behaviour.

\begin{figure*}
 \centering
  \begin{subfigure}{.45\textwidth}
   \includegraphics[width=0.93\textwidth]{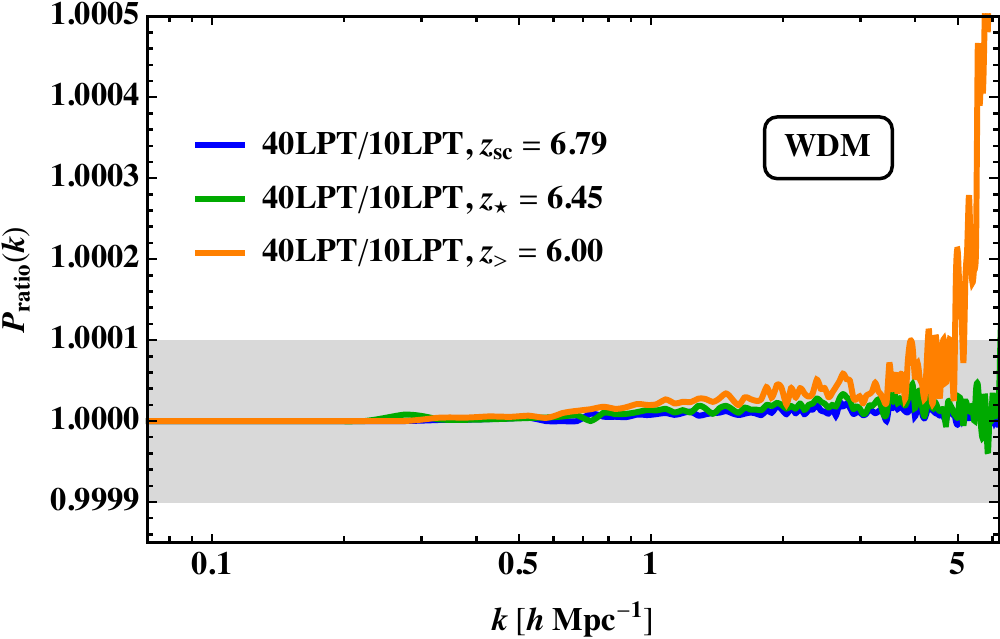}
  \end{subfigure}%
  \begin{subfigure}{0.45\textwidth}
   \includegraphics[width=0.9\textwidth]{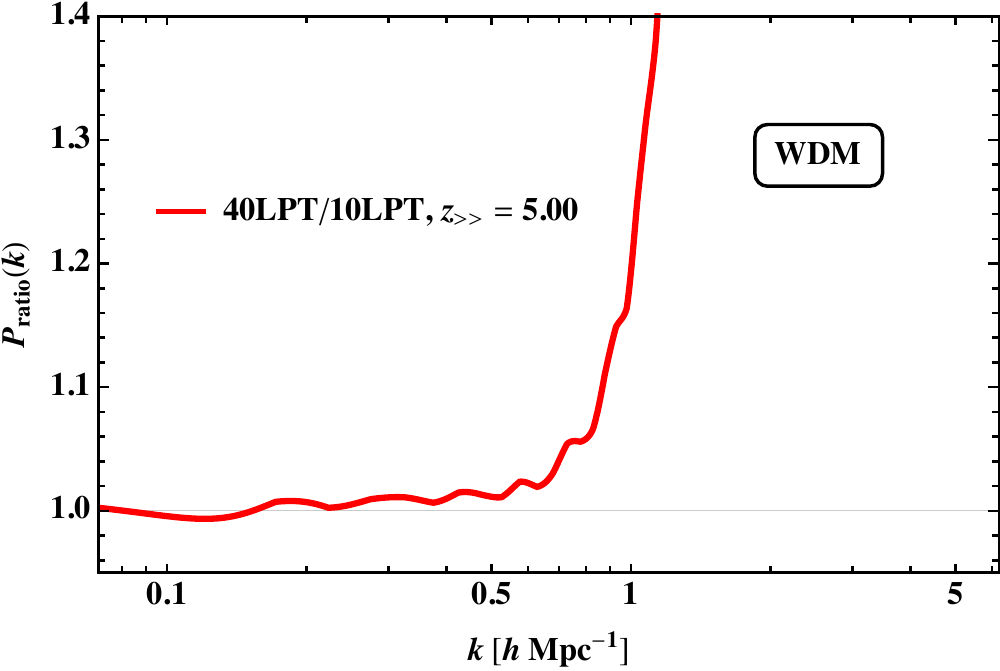}
  \end{subfigure}%
   \caption{Ratio of power spectra of the truncated Jacobian $J^{\{n\}}$ for $n=40$ versus $n=10$ at various redshifts for  $m_{\rm wdm} = 0.25$\,keV and a resolution of $N=256^3$ with $L_{\rm box} = 125 \Mpch$. At late redshifts ($z < z_\star$), 40LPT exemplifies stronger divergent behaviour than that observed at 15LPT (see Fig.\,\ref{fig:ratioJlatetime}).}
  \label{fig:ratioPJ40lpt}
\end{figure*}

Finally, in Fig.\,\ref{fig:ratioPJ40lpt} we show for our base WDM cosmology  ratios of powerspectra of the truncated Jacobian $J^{\{n\}}$ for $n=40$ versus $n=10$ (i.e., the same plot as in the lower panel of Fig.\,\ref{fig:ratioJlatetime} but at $n=40$ instead $n=15$). At shell-crossing ($z_{\rm sc}= 6.79$, blue line) and even at the estimated redshift of convergence ($z_\star=6.45$, green line) the ratios conform to unity to very high precision, which is another indication that LPT is essentially fully converged already at $n=10$ at those redshifts. Beyond the redshift of convergence however ($z_> =6$ in orange and $z_{\gg}=5$ in red), LPT clearly exhibits divergent behaviour, which is much stronger than observed in Fig.\,\ref{fig:ratioJlatetime}, and an  expected feature on theory grounds.

\section{Searching for the first shell-crossing in time and space}\label{app:zsc-results}

Here we provide more details on how to determine the time of first shell-crossing as well as possible refinement levels.

\paragraph*{Numerical methodology.} As discussed in the main text, in Lagrangian coordinates the signature of shell-crossing is the first vanishing of the Jacobian, i.e.,
\be
 J(\fett{q}_{\rm sc}, z_{\rm sc} )= \det \left( \mathbb{1} + \nabL \fett{\psi}(\fett{q}_{\rm sc}, z_{\rm sc}) \right) \stackrel ! = 0 \,,
\ee
at shell-crossing location $\fett{q} = \fett{q}_{\rm sc}$ and corresponding redshift $z= z_{\rm sc}$.
In numerical implementations, searching for an exact zero is not feasible, instead one chooses a small but positive cutoff $\epsilon>0$ and searches in a respective interval. Furthermore,
due to the random nature of the initial conditions, field computations are performed at collocation points $q_{i,j,k}$ where $i,j,k$ label the Lagrangian grid points in the three space dimensions.
Thus, the search of the first shell-crossing in numerical implementations amounts to searching for
the earliest $z_{\rm sc}$ for which
\be
 0 < J  (q_{i_{\rm sc},j_{\rm sc}, k_{\rm sc}}, z_{\rm sc}) < \epsilon 
\ee
is satisfied for the first time, where $q_{i_{\rm sc},j_{\rm sc}, k_{\rm sc}}$ is the corresponding collocation point (see further below for the study of shell-crossings at off-grid locations). 
In this work we choose a cutoff of $\epsilon = 10^{-3}$, which by virtue of the  relation $\delta= 1/J-1$, amounts to resolving overdensities of $\delta  \gtrsim 10^3$.

Higher peaks in the density can be easily resolved if desired, namely by noticing that close to shell-crossing, the redshift dependence on the shell-crossing redshift is roughly linear. Thus, by applying simple linear extra- or interpolations we can easily reach shell-crossings to a cutoff precision of $\epsilon = 10^{-9}$ in the Jacobian or even more -- up to machine precision.

\paragraph*{Theoretical convergence of shell-crossing time.} For the above methodology one of course needs to fix {\it a priori} the perturbative truncation order for the Jacobian. For computational efficiency, it is constructive to begin the search at around LPT order $n=3$, which is in most cases already sufficient to determine the grid position of the shell-crossing. We note however, that in very rare cases, we observed a change of shell-crossing location at around orders 3-5. Nonetheless, as soon as the shell-crossing redshift is converged to one decimal precision, which is usually achieved around order 6, we find that the spatial location of the first shell-crossing does not change anymore (and even converges fast to sub-grid locations; see below).

To get an intuition of the convergence speed, we provide in the following some typical examples for given realization and settings. For a standard $\Lambda$CDM cosmology with $N=256^3$ and $L_{\rm box} = 125\Mpch$, the $n$th-order estimate on $z_{\rm sc}$, starting from order $n=3$ to higher orders is
\begin{align}
  z_{\rm sc} =\, &14.02 , \,  14.25 , \, 14.37  , \, 14.44  , \, 14.48 , \, 14.51 , \, 14.52 , \nonumber  \\
  &14.53 , \, 14.54 , \, 14.54 ,\,14.54,
\end{align}
and thus, in the present case converges to two decimals at orders $n \geq 11$. For the same resolution settings but for our reference WDM model with $m_{\rm wdm} =0.25$\,keV, by contrast, we typically find the trends
\begin{align}
  z_{\rm sc} =\, & 6.26 , \, 6.47 , \, 6.60 , \, 6.67 , \, 6.71 , \, 6.72 , \, 6.75 , \nonumber \\ 
    &  6.76 , \, 6.77 , \, 6.78 , \, 6.79 , \, 6.79 , \,6.79 \,,
\end{align}
and thus converges here for orders $n \geq 13$.

\paragraph*{Sub-grid convergence of shell-crossing locations.} In the present paper we set our shell-crossing criterion on grid points. Although possibly debatable as regards to the physical usefulness -- since UV effects may or may not play a role -- one can however also search for shell-crossings at off-grid locations.
From the side of numerical and theoretical convergence, however, we find such avenues to be helpful and thus, we provide in the following some details.

While LPT is based on a continuous fluid description, our numerical implementation thereof (involving fast Fourier transforms) evaluates continuous fields, such as the Jacobian, only at collocation points. However, the Jacobian can be easily evaluated at off-grid locations using interpolation methods. We have done so, adopting a cubic-spline interpolation.

Na\"ively, searching for off-grid shell-crossings could be performed by first searching for the minimum in the Jacobian  at grid locations, and then searching for the local minimum near those grid points. However, we find that this somewhat simplistic procedure occasionally misses the true global minimum, therefore for this case we have  implemented a more sophisticated approach, explained below.

As explained in the main text, the seeds/shapes that shell-cross first always arise from the tail in the statistical sample. Specifically, we find that initial densities (say, at $z_{\rm ini}=100$) at Lagrangian shell-crossing locations are at least 3$\sigma$ deviations in comparison to the whole sample. Another criteria of those seeds is that the $L^2$ norm of the Hessian of the local gravitational potential, $\| \nabL \nabL \varphi(z_{\rm ini})\|$, is always very large: for all realizations considered we find these norms to be at least  4$\sigma$ deviations at shell-crossing locations (sometimes even 7$\sigma$).  Actually, there is a good theoretical reason why $\| \nabL \nabL \varphi(z_{\rm ini})\|$ is relevant for detecting shell-crossings, since it essentially measures the strength of the initial velocity gradients \citep[see e.g.,][]{Rampf:2015mza,Michaux:2020}.

With this knowledge we can now introduce our refined procedure of finding the off-grid shell-crossing locations.
\begin{enumerate} 
 \item  At high redshift $z_{\rm ini}$ we search for locations on the grid where the Lagrangian density and  $\| \nabL \nabL \varphi(z_{\rm ini})\|$ are extraordinarily large, respectively at least 3 and 4$\sigma$ deviations from their mean. 
 \item This search ends up with a list of Lagrangian grid positions where shell-crossing might occur (for $N=256^3$, usually around 50 grid positions). 
 \item Finally we employ standard numerical minimization procedures to search for the local minima of the Jacobian near those grid locations.
\end{enumerate}
The first shell-crossing is the most minimal value of those local minimas.
Note that for random initial conditions the first shell-crossing always occurs off grid, which consequently leads to a slight shift of shell-crossing redshifts to earlier times than reported above.

\begin{table}
\begin{center}
\begin{tabular}{ccccc}
 LPT  order & $z_{\rm sc}$ & $q_{x}^{\rm sc}$  & $q_{y}^{\rm sc}$   & $q_{z}^{\rm sc}$  \\
\hline
  3   & 18.95  &  42.1307  &  163.342  &  64.5681 \\
  4   & 19.22  &  42.1281  &  163.344  &  64.5664 \\
  5   & 19.35  &  42.1264  &  163.344  &  64.5658 \\
  6   & 19.40  &  42.1257  &  163.344  &  64.5655 \\ 
  7   & 19.42  &  42.1253  &  163.344  &  64.5653 \\
  8   & 19.43  &  42.1251  &  163.344  &  64.5652 \\
\hline
\end{tabular}
\end{center}
\caption{Theoretical convergence of the shell-crossing locations, for a random standard $\Lambda$CDM cosmology with $N=256^3$ and $L_{\rm box} = 125\Mpch$. Values $q_i^{\rm sc}$ denote the dimensionless positions between grid points.}
\label{tab:offgrid}
\end{table}

Finally, we demonstrate the subgrid convergence on a case example for a standard $\Lambda$CDM cosmology with $N=256^3$ and $L_{\rm box} = 125 \Mpch$.  The results, as a function of LPT orders are summarized in Table~\ref{tab:offgrid}, with  $q_{x,y,z}^{\rm sc}$ denoting the perturbative estimates of the dimensionless coordinate inter-grid values of first shell-crossing. Clearly, while convergence in location is observed in all coordinate directions, the speed of convergence varies, which is due to the local forcing that acts on the collapsing seeds.

\section{Study of outliers in the Domb--Sykes plot} \label{app:outliers}

\begin{figure*}
\centering
\begin{subfigure}{.33\textwidth}
  \centering\captionsetup{width=.85\linewidth}%
  \includegraphics[width=.95\linewidth]{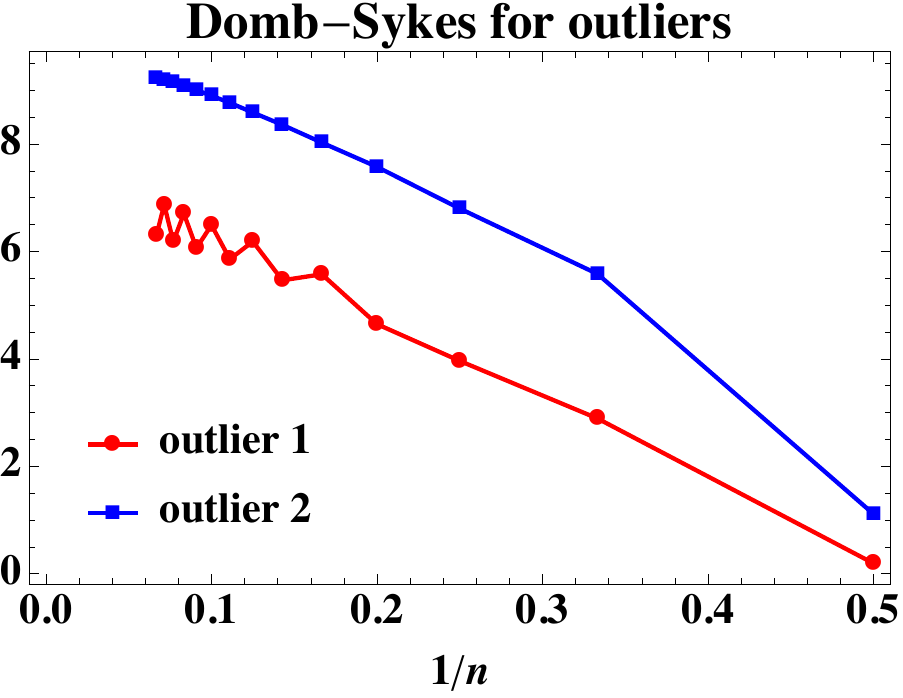}
  \caption{Domb--Sykes plot for two random realizations at shell-crossing locations, showing oscillatory pattern (``outlier 1'', red lines) and a bend (``outlier 2'', blue lines). }
  \label{fig:outsub1}
\end{subfigure}%
\begin{subfigure}{.33\textwidth}
  \centering\captionsetup{width=.8\linewidth}%
  \includegraphics[width=.95\linewidth]{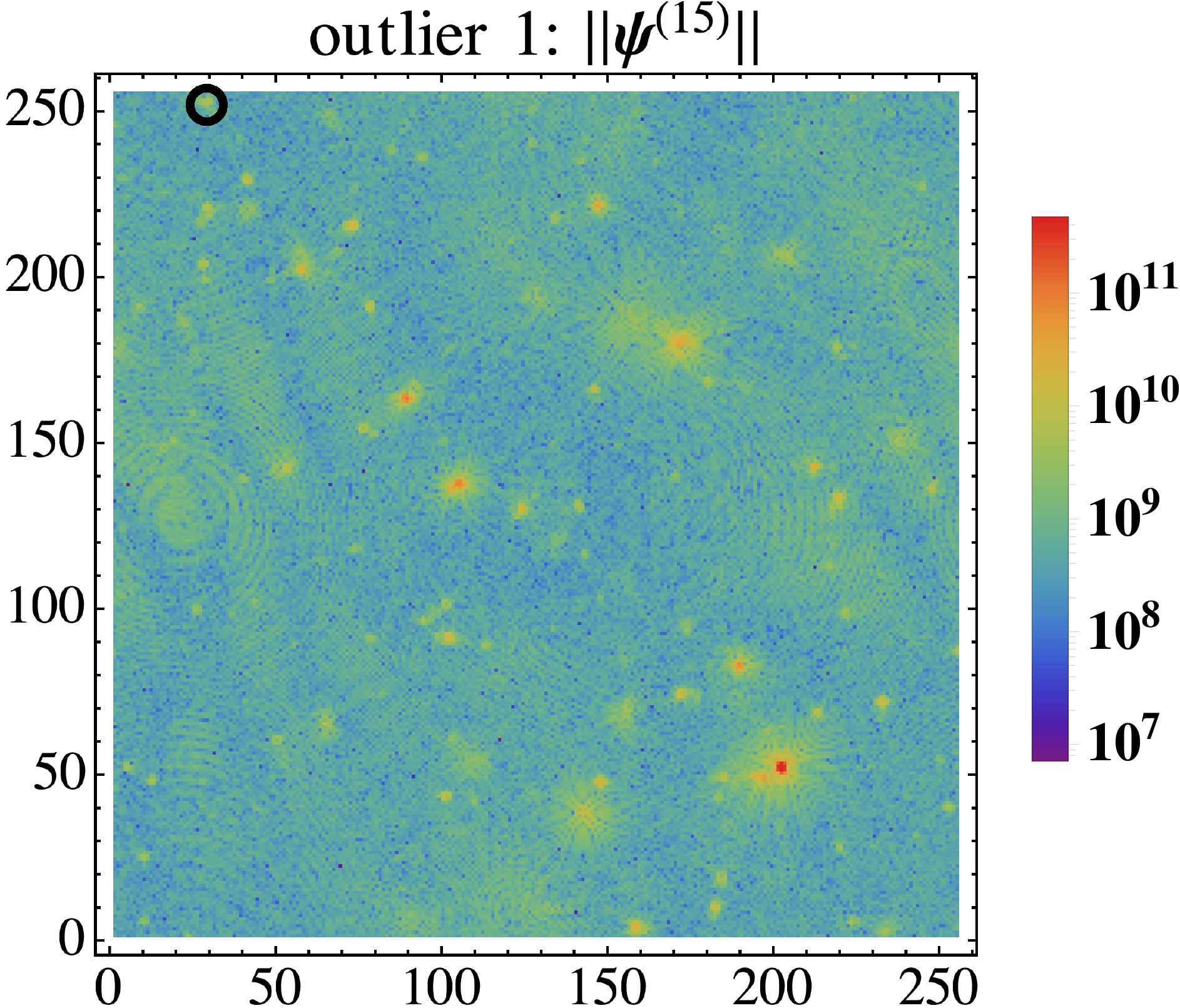}
  \caption{Slice of real-space realization of the displacement norm of ``outlier 1'' at 15th order. The first shell-crossing location is denoted by a black circle.}
  \label{fig:outsub2}
\end{subfigure}
\begin{subfigure}{.33\textwidth}
  \centering\captionsetup{width=.8\linewidth}%
  \includegraphics[width=.95\linewidth]{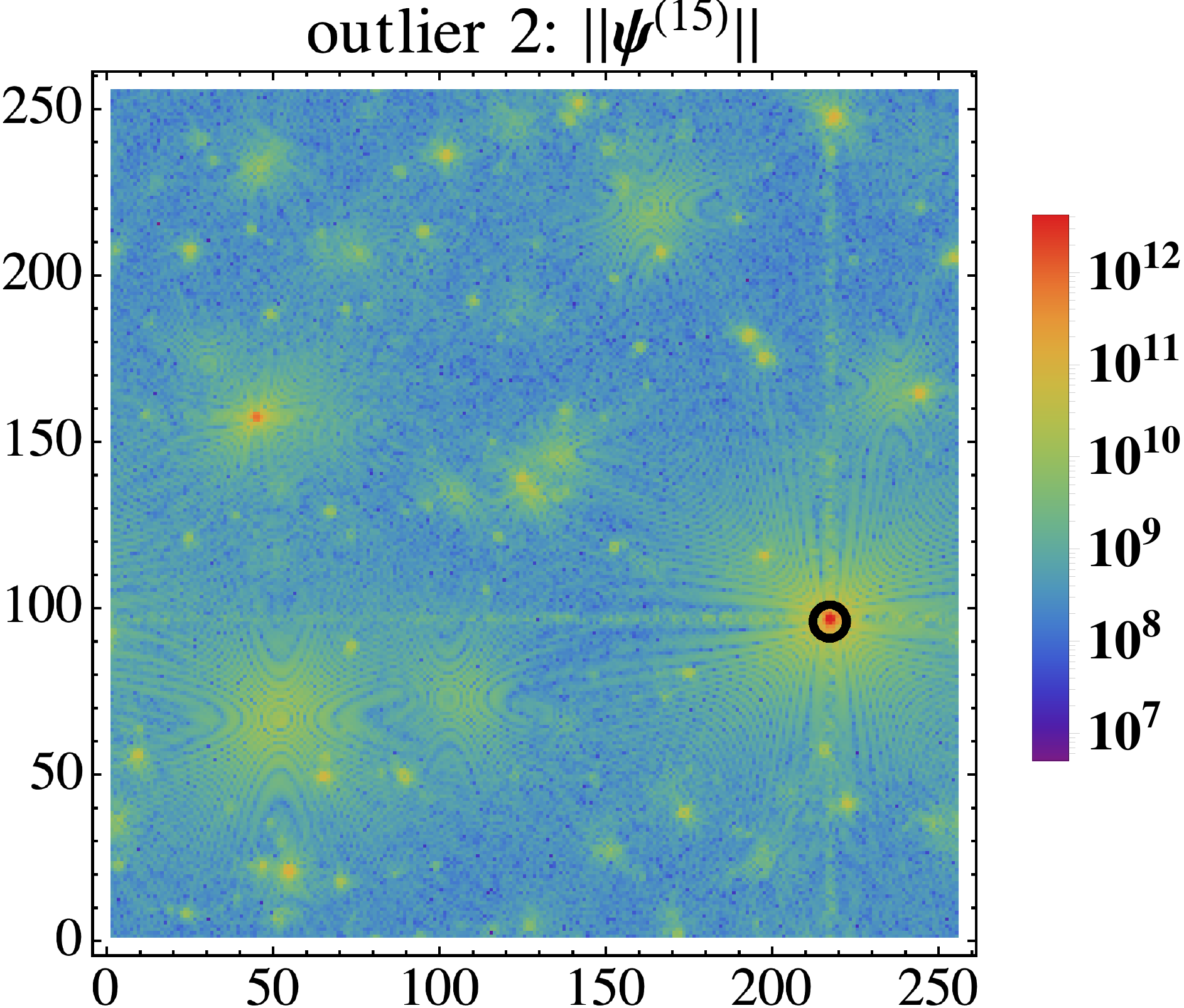}
  \caption{Same as Fig.\,\ref{fig:outsub2} but for ``outlier 2''. Note the arc-like artefacts that can appear at very large Taylor orders due to the Gibbs phenomenon.}
  \label{fig:outsub3}
\end{subfigure}
\caption{Study of outliers that emerge occasionally when investigating the LPT convergence with random initial conditions.}
\label{fig:outliers}
\end{figure*}

For reasons of completeness,  we investigate here in details the outliers related to the employed convergence tests, which are generally expected to occur due to the random nature of the initial conditions \citep[see e.g.][]{PODVIGINA2016320,Michaux:2020}. As we shall see,  for the detected outliers,  one can still estimate the radius of convergence fairly accurately, however in some cases the nature of the  convergence-limiting singularities remains shrouded, at least within the currently implemented methods.

We have observed two types of outliers, called ``outlier 1'' and ``outlier 2'' in the following, with peculiar characteristics as summarized in Figs.\,\ref{fig:outliers}, where  the leftmost panel shows the Domb--Sykes plot, while the central and rightmost panels show snapshots of real-space realizations of the norm of the displacement coefficients at order $n=15$ (the effects that we investigate here are vastly suppressed and thus hardly visible at lower orders). 
Both outliers were obtained from standard $\Lambda$CDM realizations with $N=256^3$ grid points and corresponding box length $L_{\rm box} = 125 \Mpch$. 

For outlier 1, the collapsing region has at $z_{\rm ini}=100$ the eigenvalues of the Jacobian of  $\lambda_1^{\rm ini} =0.862$, $\lambda_2^{\rm ini}= 0.971$ and $\lambda_3^{\rm ini} = 0.990$ and thus has a fairly quasi-one-dimensional shape which is actually well maintained during collapse for the larger $\lambda_i$'s ($\lambda^{\rm sc}_1 = 0.000$, $\lambda^{\rm sc}_2 = 0.781$ and $\lambda^{\rm sc}_3 = 0.933$). 
One prominent feature of outlier 1 is the oscillating behaviour in the Domb--Sykes plot at large Taylor orders (red line in Fig.\,\ref{fig:outsub1}). As discussed by  \cite{vanDyke1974} for non-cosmological problems, such oscillatory features are either caused by singularities that are not on the real time axes but in the complex plane, or due to an overlap of adjacent singularities. In the present case, the former is ruled out since the signs of the Taylor coefficients are not changing. Nonetheless we have verified the absence of complex singularities  by employing the estimator of \cite{MercerRoberts1990}, which, as anticipated, ended up with a null result. Therefore, we conclude that outlier 1 is due to overlapping singularities. Despite these oscillations, we find that the extrapolation in the Domb--Sykes plot to $1/n \to 0$ still reveals reliable estimates on the radius of convergence, but, as noted above, the nature of the singularities remains unclear in such cases.

For outlier 2, the collapsing Lagrangian region has the initial eigenvalues of  $\lambda_1^{\rm ini} =0.884$, $\lambda_2^{\rm ini}= 0.928$ and $\lambda_3^{\rm ini} = 0.932$ which points to a fairly spherical shape. Its Domb--Sykes plot (blue lines in Fig.\,\ref{fig:outsub1}) shows an almost linear relation, however with a slight bend  at large Taylor orders. This slight bend appears to be not of physical origin  but is caused by the Gibbs phenomenon, i.e., by reverting sharp features in $\| \fett{\psi}^{(n)} \|$ from Fourier space to real space. This phenomenon manifests as arc-like artefacts in Fig.\,\ref{fig:outsub3} which usually tend to be strong near extremal shell-crossing locations (marked by black circle in Fig.\,\ref{fig:outsub3}).
As mentioned in the main text, while for standard $\Lambda$CDM realizations, the appearance of the Gibbs phenomenon usually affects the Domb--Sykes plot at orders beyond $n=15$, in the case of outlier 2 these artefacts already appear at order $n\simeq 10$. We believe this to originate from the fairly quasi-spherical nature of the collapsing seeds for outlier 2, which leads to a slower decay of amplitude in the displacement coefficients at large orders (e.g., as compared to the quasi-1D collapse shown in Fig.\,\ref{fig:outsub2}), and thereby to more pronounced sharp features in $\| \fett{\psi}^{(n)} \|$ for $n \gtrsim 10$ and thus, to the stronger appearance of the Gibbs phenomenon.

As mentioned in appendix~\ref{app:corner}, we have already implemented measures to reduce the impact of  the Gibbs phenomenon, namely by applying a top-hat filter in Fourier space before Fourier transforming the fields to real space. We speculate that this artefact could be further suppressed (i.e., pushed to even higher LPT orders) by applying optimized filters; we keep such analysis for future work. Nevertheless,  for outliers of the second type, the radius of convergence as well as the nature of the singularity can be reasonably well estimated by resorting to sufficiently low perturbative estimates in the Domb--Sykes plot.

\section{Alternative convergence test}\label{sec:pod}

While drawing the Domb--Sykes plot (see above and in the main text) is a reliable method to determine the radius of convergence, it is not always the most efficient way. An improved method is a method that makes use of the knowledge of large-$n$ asymptotic expansion of the Taylor coefficients \citep{PODVIGINA2016320}: Suppose we want to determine the radius of convergence $R$ of
\be
   f(D) = \sum_{s=1}^\infty f_s D^s \,.
\ee 
Furthermore let us assume that all coefficients are positive (as in our case when we take the norm), and that there is only a single singularity; then it follows that this single singularity must be on the real positive axis $D = D_\star$. The simplest case is to assume that the singularity has either a pole or a power-law branch point which both can be characterised by the asymptotic behaviour  $\propto (D-D_\star)^{\,\rho}$ where the exponent is a real number. It then can be shown that the large-$s$ behaviour of Taylor coefficients behaves as
\be
  f_s \simeq \gamma \,s^\alpha {\rm e}^{\,\beta s}  \,, \qquad (\text{for~} s \to \infty)
\ee 
where $\alpha = -\rho-1$ and $\beta = - \ln R$. Likewisely, we have
\be
   F_s = \ln \gamma + \alpha  \ln s + \beta s 
\ee
for the logarithms of the coefficients $F_s \equiv \ln f_s$.
Then, we perform a least-square fit of the logarithms of the displacement coefficients at the first shell-crossing point.  As a demonstrative example, let us estimate the asymptotic behaviour for a CDM cosmology with $N=256^3$ and a WDM cosmology with $N=128^3$, and  average over five realizations each.
Then,
 the least square fit for $F_s$  
reveal asymptotic behaviours of the form
\begin{align}
  &f_s^{\rm cdm} = 0.085 \, s^{-1.541} \, {\rm e}^{2.196s} \,, \qquad f_s^{\rm wdm} = 0.478 \, s^{-1.850} \, {\rm e}^{1.687s}  \,, \nonumber   
\end{align}
from which we can read off the inverse of the radius of convergence $1/R = {\rm e}^{\,\beta} = 8.990, 5.404$ respectively, 
which agree excellently with the estimates obtained via the Domb--Sykes method. 
Furthermore, the above analysis also reveals the asymptotic behaviour of the large-$n$ Taylor coefficients,
namely $(D- D_\star)^{0.541}$ for CDM and $(D- D_\star)^{0.850}$ for WDM, indicating that the first time derivative of the displacement (a.k.a.\ the velocity) blows up once convergence is lost. We remark that these results vary slightly, depending on the chosen LPT orders used for the asymptotic regime. For example,  in the WDM case, if we use for the LPT orders from $n=1$ to $40$, we have $\rho = 0.845$; if we use instead $n=20$ to $40$ we have $\rho = 0.887$.

\section{Further results to the first collapsing structures}\label{app:eandp}

\begin{figure}
	\centering
	\includegraphics[width=0.8\columnwidth]{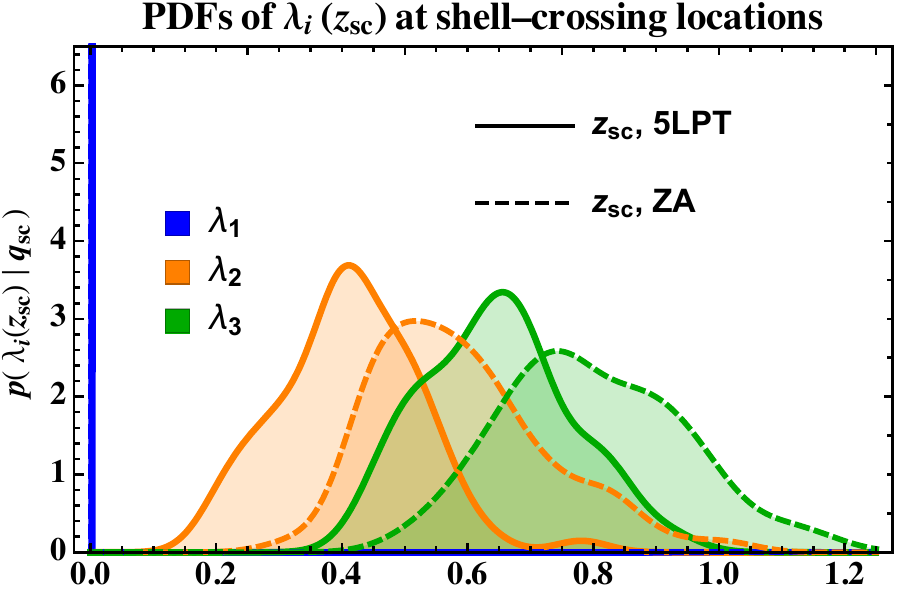}
	\caption{Conditional distributions of eigenvalues of the Jacobian at Lagrangian locations that shell-cross first, evaluated at $z_{\rm sc}$. Solid lines denote non-linear predictions according to 5th order in LPT, while dashed lines are obtained using the Zel'dovich approximation (1LPT). }
	\label{fig:PDF-ZA}
\end{figure}

Here we provide more details related to section~\ref{sec:collapsestructures}, where we investigated the evolution of the eigenvalues $\lambda_i$ of the Jacobian until shell-crossing. While the results in the main text employed fully nonlinear predictions in LPT, here we ask the question how much the leading-order LPT solution, commonly called the Zel'dovich approximation, performs in comparison. 
In Fig.\,\ref{fig:PDF-ZA} we show the conditional PDFs of the $\lambda_i$'s  that shell-cross first, evaluated at shell-crossing time. Solid lines denote the nonlinear prediction at 5LPT, while dashed lines are obtained using the Zel'dovich approximation.
While in both cases, the distribution of the eigenvalues is non-Gaussian, we observe a substantial difference in the predictions, from which we conclude that the Zel'dovich approximation is insufficient for predicting collapse features, close to the collapse, accurately.

Further we investigate the ellipticity $e$ and prolateness $p$ which are
defined with
\be
  e = \frac{\lambda_{1, \rm d} - \lambda_{3, \rm d}}{2 \sum_i \lambda_{i, \rm d}} \,, \qquad \quad  p = \frac{\lambda_{1, \rm d} - 2 \lambda_{2, \rm d} + \lambda_{3,\rm d}}{2 \sum_i \lambda_{i, \rm d}} 
\ee
\citep[see e.g.][]{Bardeen1986}, where the $\lambda_{i, \rm d}$'s are the three real eigenvalues of the  symmetrized deformation tensor $\nabL\fett{\psi}$ and relates to the eigenvalues of the Jacobian matrix according to
 $\lambda_{i, \rm d} = \lambda_i -1$.

\begin{figure}
	\centering
	\includegraphics[width=0.8\columnwidth]{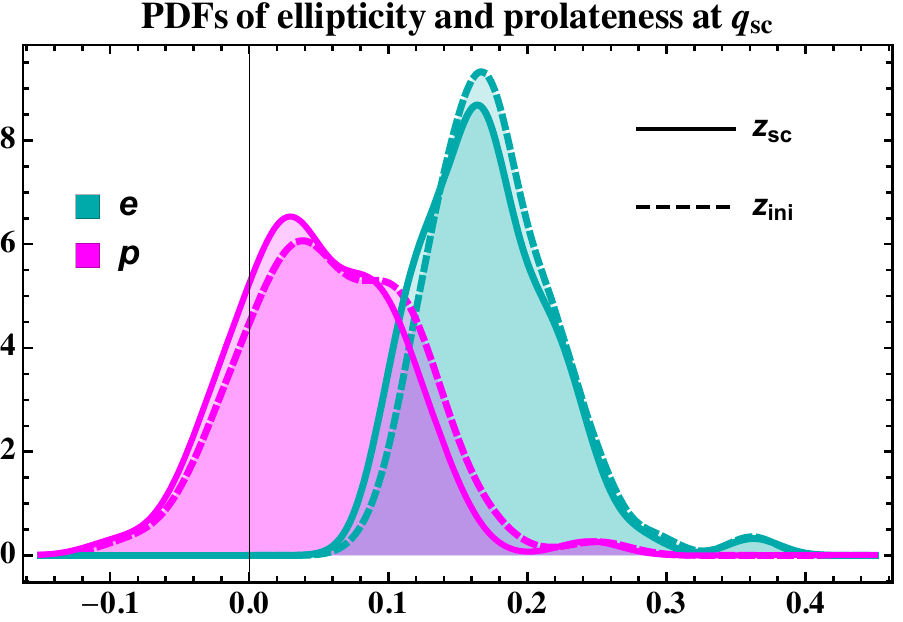}
	\caption{PDFs of $e$ and $p$ at the first shell-crossing location, obtained from 65 random realizations. Solid lines denote shell-crossing times, dashed lines the  distributions at $z_{\rm ini} =100$ evaluated at the same Lagrangian positions.  }
	\label{fig:PDFeandp}
\end{figure}

As discussed in section~\ref{sec:collapsestructures}, we begin searching for the first shell-crossing location for given random initial conditions, and then evaluate $e$ and $p$ for these locations at the initial redshift $z_{\rm ini}= 100$ and shell-crossing redshift $z_{\rm sc}$. The resulting conditional PDF is shown in Fig.\,\ref{fig:PDFeandp}. Evidently, both distributions hardly change during the course of evolution, except with a slight tendency to approach zero, i.e., collapsing objects tend to  become more spherical.

\begin{figure}
	\centering
	\includegraphics[width=0.85\columnwidth]{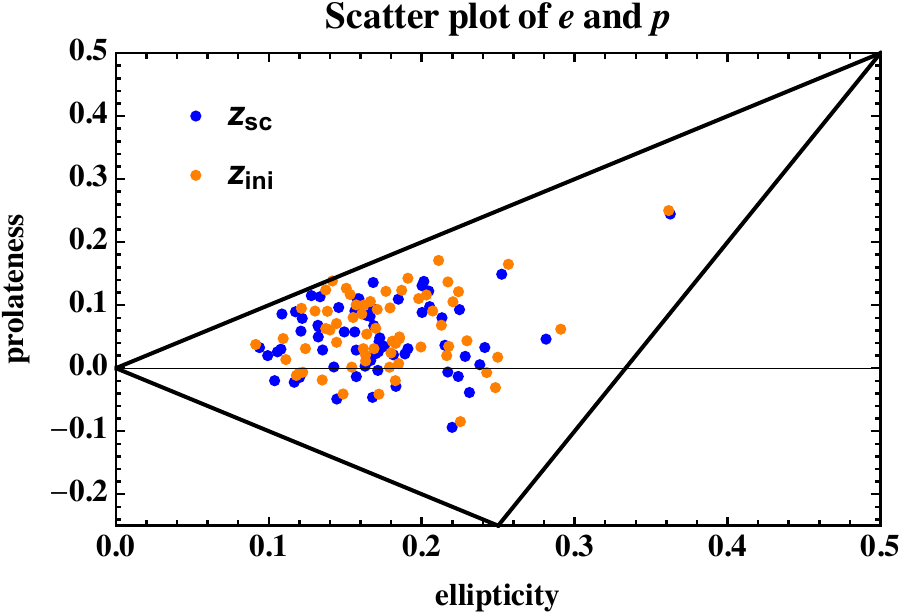}
	\caption{Scatter plot of pairs of $(e,p)$ of Lagrangian seeds that will shell-cross first, at $z_{\rm ini}=100$ (orange dots) and at $z_{\rm sc}$ (blue dots). For extreme peaks using Gaussian random initial conditions (and to linear order), those pairs should lie within the marked triangle.}
	\label{fig:scatter}
\end{figure}

Finally we show a scatter plot of all pairs $(e,p)$ in Fig.\,\ref{fig:scatter}. 
For comparison we also draw the triangle as in \cite{Bardeen1986} (see their Fig.\,7, obtained from linear considerations);  our results indicate a bias towards a positive prolateness compared to their peaks with comparable height.

\vfill

\bsp	
\label{lastpage}
\end{document}